\documentclass[twocolumn,epjc3]{svjour3}
\usepackage{graphicx}
\usepackage{xcolor}
\newcommand{\newchangecolor}{blue}
\DeclareRobustCommand{\newchange}[1]{{\color{\newchangecolor}#1}}

\DeclareRobustCommand{\rev}[1]{{#1}}
\newenvironment{revision}{\par\begingroup}{\par\endgroup}
\usepackage{cite}
\usepackage{hyperref}
\usepackage[T1]{fontenc}
\usepackage[utf8]{inputenc}
\usepackage{amsmath,amssymb,booktabs}
\usepackage{lmodern}
\usepackage{microtype}
\begin{document}
\title{Charged relativistic quark-matter spheres: exact solutions and analytical admissibility bounds}

\author{D. Santana
\thanksref{e3,addr3}
        %\and
        %C. Usca\thanksref{e4,addr3}
        \and
        J. Andrade\thanksref{e5,addr3}
        \and
        Daniela Tatiana Castañeda Ortiz\thanksref{e8,addr3}
        \and
        B. Yanchapanta\thanksref{addr4}
        \and
        K. Mindiola\thanksref{addr4}
}
\thankstext{e3}{e-mail:
\href{mailto:diego.santana@espoch.edu.ec}{\nolinkurl{diego.santana@espoch.edu.ec}}}
%\thankstext{e4}{e-mail:
%\href{mailto:icrisus25@gmail.com}{\nolinkurl{icrisus25@gmail.com}}}
\thankstext{e5}{e-mail:
\href{mailto:julio.andrade@espoch.edu.ec}{\nolinkurl{julio.andrade@espoch.edu.ec}}}
\thankstext{e8}{e-mail:
\href{mailto:tatiana.castanieda@espoch.edu.ec}{\nolinkurl{tatiana.castanieda@espoch.edu.ec}}}

\institute{Escuela Superior Politécnica de Chimborazo, (ESPOCH), Riobamba 060155, Ecuador\hypertarget{addr3}{}\label{addr3}
\and
Independent researcher \hypertarget{addr4}{}\label{addr4}}

\titlerunning{\newchange{Charged quark-matter spheres: admissibility bounds}}
\maketitle
\begin{abstract}
We investigate source compatibility and the static physical admissibility of charged anisotropic quark-matter spheres in the Einstein--Maxwell framework. We first show that a purely electromagnetic extension of a regular neutral interior cannot be generated through a purely radial minimal geometric deformation when the temporal metric potential and material variables remain fixed: the Maxwell stress identities and central regularity force both the deformation and the electric field to vanish. We then construct a family of exact solutions by retaining the Tolman-IV temporal metric potential, imposing an effective radial MIT-bag-type relation with a constant bag parameter, and prescribing a regular cubic charge distribution. Matching to a Reissner--Nordstr\"om exterior determines the dimensionless bag parameter and relates charge to compactness. Nonnegative tangential surface pressure yields an explicit upper bound on the charge parameter. Using Bernstein polynomials, we prove that this bound is necessary and sufficient for static admissibility, with $y = (R/a)^{2}\geq y_{0}\simeq 0.64743$. At $R=a$, it gives $0\leq\epsilon\leq 21/2665$. We also derive a charge bound from the Schwinger field scale. For a given surface-to-scale ratio, the charged members are more compact than the neutral one. The maximum charge is reached when the material tangential pressure vanishes at the surface. The certified intervals describe static equilibria; the question of dynamical stability is left open.
\end{abstract}

\section{Introduction}

The composition of matter deep inside the most compact stars remains uncertain. At the densities of interest, deconfinement may occur and matter may contain up ($u$), down ($d$), and strange ($s$) quarks. According to the Bodmer--Witten hypothesis, bulk strange quark matter at zero pressure could be energetically favoured over ordinary nuclear matter on a per-baryon basis. Its absolute stability has yet to be established \cite{bodmer,witten,farhi}. If the hypothesis holds, quark matter need not be restricted to the core of a neutron star: it could constitute a self-bound star of the kind studied by Alcock, Farhi, and Olinto \cite{alcock1986strange,weber2005strange}.

Unlike matter whose density tends to zero at the stellar boundary, self-bound matter may have a finite density at zero pressure. The mass and radius of the star must nevertheless be determined from gravitational equilibrium. We describe quark matter using the MIT bag model \cite{chodos,farhi}. For cold, massless quarks with no interactions, the equation of state takes the form $p=(\rho-4\mathcal{B})/3$, where $\mathcal{B}$ is the bag constant. The zero-pressure density is then $\rho=4\mathcal{B}$. We adopt this relation for its analytical simplicity, bearing in mind that it neglects the strange-quark mass and interactions between quarks \cite{farhi,weber2005strange}. The use of this equation of state does not establish the absolute stability of strange matter.

More detailed descriptions of quark matter are available from quantum chromodynamics (QCD). Perturbative QCD includes effects of the strong interaction and the strange-quark mass, although its predictions become less certain when applied to the densities expected inside compact stars \cite{kurkela2010cold,fraga2014interacting}. Other models allow the bag constant to vary with density or consider quark pairing in color-superconducting phases \cite{alford2008color,sedaghat2025constraint}. These effects can change the relation between pressure and energy density. Our choice of a constant bag parameter, with $dp_r/d\rho=1/3$, is therefore a simplification that allows a clear analytical treatment. It should not be taken as a general prediction of QCD.

Here we are interested in stars composed predominantly of self-bound quark matter, rather than hybrid stars in which a quark core is surrounded by hadronic matter. Evidence favouring quark cores in massive neutron stars does not establish the existence of the former class of objects \cite{annala2020evidence}. Nor can the equilibrium solutions derived below determine the composition of an observed star.

To build a self-gravitating sphere, we combine the bag equation of state with the Einstein--Maxwell equations and boundary conditions. The bag relation is imposed on the \emph{radial material pressure} $p_r$, while the tangential pressure $p_t$ follows from the field equations. We also prescribe a nonzero electric charge distribution to study how charge affects equilibrium. Bulk quark matter is ordinarily treated as nearly charge neutral. The charge introduced here is a macroscopic input, since the bag model offers no mechanism for its generation or retention. We keep the electromagnetic contribution separate from the material density and pressures.

For an anisotropic fluid, hydrostatic equilibrium involves the difference between the tangential and radial pressures in addition to the radial pressure gradient. The work of Bowers and Liang demonstrated that this contribution can alter the equilibrium mass and surface redshift of relativistic spheres \cite{bowers}. Exact anisotropic solutions subsequently provided settings in which these effects could be followed throughout the interior \cite{maurya2015,shee2018,maharaj2014some,sunzu2014charged,bayin1982anisotropic, dev2002anisotropic, mak2003anisotropic, chaisi2006anisotropic, herrera2008all,naseer2024estimating,campos2026,cosenza1981some}. For example, Maurya et al. constructed anisotropic compact-star models based on a deformation of the Durgapal--Fuloria geometry \cite{maurya2015}, while Shee et al. combined a Tolman-V potential with a bag equation of state in a strange-star model \cite{shee2018}. A radial equation of state alone leaves the tangential pressure undetermined. To close the static equations one must prescribe another relation or fix part of the geometry; a perturbative treatment also requires a law for the response of the tangential stress. In a charged star the electric field enters both the gravitational field equations and the hydrostatic balance. Bekenstein's treatment of relativistic charged fluid spheres provides a basis for studying these effects consistently within the Einstein--Maxwell equations \cite{bekenstein}. For anisotropic strange matter, Deb et al. employed the MIT bag relation together with a cubic charge profile and matching to a Reissner--Nordstr\"om exterior \cite{deb2018}. Related charged bag-model solutions have been developed with Tolman--Kuchowicz geometry by Maurya and Tello-Ortiz \cite{maurya}, and with a Tolman-IV potential and a density-dependent bag parameter by Goswami et al. \cite{goswami}. These studies provide close points of comparison for the present problem. They also make it necessary to distinguish the prescribed charge distribution, the material anisotropy and the electromagnetic stress when assessing the behaviour of a solution.

The sign of the pressure anisotropy also requires a physical interpretation. As noted by Campos et al. \cite{campos2026}, phase transitions, solid or crystalline phases, condensates, and intense magnetic fields are potential scenarios for the emergence of directional stresses in compact matter. A negative anisotropy, $\Delta = p_t - p_r < 0$, describes a regime in which the radial pressure exceeds the tangential pressure. Its sign depends on the constitutive response and—where applicable—on the orientation of the deformation or the field; the mere presence of a phase transition or a condensate does not, by itself, determine this sign. Elastic stresses within a deformed solid provide a concrete mechanism that explains the inequality between the principal pressures. In particular, quark matter in a crystalline color-superconducting state possesses a non-zero shear modulus and is capable of sustaining such stresses \cite{mannarelli2007rigidity}.

Phenomenological studies of quark stars consider both the dominance of radial pressure and that of tangential pressure, including models that employ an equation of state for quarks with interactions \cite{pretel2024interacting, mondal2025quasinormal}. In these models, corrections arising from interactions and quark pairing modify the radial equation of state, while an independent prescription governs the anisotropy. These results justify the possibility of considering both signs for $\Delta$ within an effective model, without the need to attribute the resulting difference in stresses to a specific microscopic mechanism.

A complementary approach to constructing anisotropic interiors is gravitational decoupling (GD) \cite{ovalle2017decoupling}. Its methodological roots lie in the minimal geometric deformation approach developed in the braneworld context, including a conjecture concerning physically relevant exact stellar solutions on the brane \cite{ovalle2008searching}. In its minimal geometric deformation formulation (MGD), a known source is supplemented by an additional sector whose effect is encoded in a deformation of the radial metric potential, while the temporal potential is retained \cite{ovalle}. The construction then requires a suitable seed and an additional condition to close the resulting equations. This method has yielded a variety of interior solutions, including those presented in \cite{gabbanelli2018gravitational,morales2018charged, torres2019anisotropic,andrade2021stellar,andrade2023anisotropic,naseer2026polytropes,andrade2022isotropic,andrade2024wyman,naseer2024complexity,maurya2025complexity,maurya2023complexity, yousaf2024energy, berkimbayev2026energy}.

This condition can incorporate material information, as illustrated by the use of a radial polytropic equation of state to integrate relativistic anisotropic Lane--Emden systems from known interior solutions \cite{santana2022}. The framework has also been extended to charged configurations, including deformations of a charged Heintzmann background \cite{morales2018charged} and studies of electromagnetic effects on stellar complexity \cite{albalahi}. These applications draw attention to the distinction between the electromagnetic field already present in the background and the additional anisotropic source introduced through the deformation. These developments show that the properties of the resulting fluid depend on both the initial geometry and the condition imposed on the deformed system.

When the temporal metric potential is also allowed to change, energy exchange between the constituent sources becomes an additional aspect of the construction. This extension of GD is based on extended minimal geometric deformation (EGD) \cite{ovalle2019decoupling}. This extension is relevant because its analytical structure allows the study of energy exchange between the relativistic fluids supporting the self-gravitating sphere described by the resulting interior solution. This method has been used to analyse energy exchange between a generic isotropic fluid and a polytrope \cite{ovalle2022energy, contreras2022energy}. Related studies using this approach include \cite{andrade2022isotropic, andrade_energy2024, andrade_exchange2025, chicaiza2025,maurya2023complexity,yousaf2024energy,berkimbayev2026energy}. More recently, the coexistence of the Einstein-universe fluid with a Tolman-IV-type source has been analysed together with the response of the resulting configuration to perturbations \cite{andrade_coexist2026}. These investigations connect the geometric deformation with the interaction between the sources and reinforce the need to ensure that each additional stress tensor satisfies the properties of the physical component it represents.

Decoupling methods have also been applied to strange-star models beyond four-dimensional general relativity, as well as to a variety of other astrophysical and gravitational contexts \cite{sharif2021compact, ovalle2018black, contreras2021gravitational, misyura2024non, naseer2025thermodynamic, maurya2022anisotropic, pradhan2025gravitationally, casadio2019isotropization, hua2026regular, casadio2024gravitational, Andrade2026}. Bag-model configurations have been investigated in five-dimensional Einstein--Gauss--Bonnet gravity and through complete geometric deformation in $f(Q)$ gravity \cite{maurya_egb,maurya_fq}. The stellar structures obtained in these theories depend on the gravitational couplings as well as the chosen deformations. Each theory has its own field equations and exterior geometry. In Einstein--Maxwell theory, we examine the restrictions imposed by the bag relation and the chosen electric field and temporal potential.

The assessment of these configurations extends beyond the construction of an equilibrium solution. Studies of decoupled interiors with Tolman-IV-type complexity have examined the appearance of cracking and overturning under perturbations \cite{andrade_cracking2022}. Radial pulsations address the evolution of small radial displacements and require a specification of the material response away from equilibrium, particularly when the pressure is anisotropic \cite{dev,horvat,pretel}. For strange stars, radial stability has been investigated in the presence of anisotropy and, separately, electric charge \cite{anisostability,chargedstability}; charged-star fluid modes have also been calculated in the Cowling approximation \cite{cowling}. This body of work provides a basis for distinguishing the conditions that constrain an admissible equilibrium background from the additional assumptions needed to determine its oscillation spectrum.

The model has several limitations. We use the simplest MIT bag relation with a constant $\mathcal{B}$ and apply it only to the radial pressure. The tangential pressure is obtained from the field equations, not from a microscopic QCD model of anisotropic matter. We also choose the cubic charge profile and the Tolman-IV temporal potential rather than deriving them from the physics of quarks or charge transport. Effects such as density-dependent bag terms, temperature, quark pairing, and a possible transition to hadronic matter are not included. Finally, the conditions verified here describe static equilibrium; they do not establish stability against radial oscillations. That question requires a model for how the anisotropic matter and electric current respond to perturbations. These limitations do not affect the exact equilibrium results, but they are important when relating them to real compact stars.

The most closely related models differ in how the field equations are closed. Deb et al. \cite{deb2018} prescribe a quadratic density profile and a cubic charge distribution, along with a radial relation based on the MIT bag model. Goswami et al. \cite{goswami} specify the temporal potential of the Tolman IV solution, the anisotropy, and the electric field, while allowing the bag parameter to vary. Banerjee \cite{banerjee2018tolman} investigates anisotropic neutral stars based on the bag model within a Tolman IV spacetime. Thirukkanesh and Ragel \cite{thirukkanesh2017charged} fix the radial metric and the electric field, deriving the temporal potential from a linear equation of state. Santana et al. \cite{santana2026radial} obtained neutral anisotropic configurations with the Tolman-IV temporal potential and a linear equation of state $p_r=\eta(\rho-\rho_s)$. Our neutral branch recovers their case $\eta=1/3$ and $\rho_s=4\mathcal{B}$. In the charged extension, the same temporal potential, a constant bag parameter and a cubic charge law determine the radial metric and the material stresses. We identify an obstruction to Maxwell-only minimal deformation and establish the full domain of static admissibility.

Against this background, the present work examines the compatibility of a charged source with a restricted minimal deformation and then constructs an equilibrium family directly from the coupled Einstein--Maxwell equations. We first ask whether the entire additional sector can be a Maxwell field when the temporal potential and neutral material source are left unchanged. We then retain the temporal dependence associated with Tolman IV \cite{tolman}, impose a radial bag relation with a constant bag parameter and a regular cubic charge profile, and reconstruct the radial geometry and tangential stress. The temporal ansatz is therefore used independently of the original isotropic Tolman-IV radial potential. The analysis combines surface matching, a survey in compactness and charge, and exact polynomial certificates for continuous admissible parameter intervals. Its purpose is to establish which equilibrium properties follow from the coupled equations and to identify the additional material and electrical assumptions needed for a subsequent radial-stability analysis.

\section{Field equations and material variables}
We consider a static, spherically symmetric distribution of electrically charged anisotropic matter. Throughout this work, we use geometrized Gaussian units, $G=c=1$, and the metric signature $(+---)$. In terms of the areal radius $r$, for which each symmetry sphere has area $4\pi r^2$, the interior line element is written as
\begin{equation}
\begin{aligned}
ds^2 &= e^{\nu(r)}dt^2-H(r)^{-1}dr^2-r^2d\Omega^2,\\
H(r) &\equiv e^{-\lambda(r)},
\end{aligned}
\end{equation}
where $d\Omega^2=d\theta^2+\sin^2\theta\,d\phi^2$. The functions $\nu(r)$ and $H(r)$ describe the temporal and radial geometry, respectively, and a prime denotes differentiation with respect to $r$. A regular static interior requires positive metric potentials, with $H(0)=1$ and a finite, nonzero value of $e^{\nu(0)}$.

The material source is described by an anisotropic fluid with mixed energy--momentum tensor
\begin{equation}
T^\mu{}_\nu
=\operatorname{diag}(\rho,-p_r,-p_t,-p_t),
\end{equation}
where $\rho$ is the material energy density and $p_r$ and $p_t$ are the principal pressures measured in the comoving frame. Spherical symmetry requires the two tangential pressures to coincide, but does not require equality between the radial and tangential stresses. Their difference,
$\Delta=p_t-p_r$, contributes to the radial force balance and can therefore modify the structure of the equilibrium configuration \cite{bowers}. At a regular centre, where no spatial direction is preferred, the material pressures must satisfy $p_r(0)=p_t(0)$.

We assume that the electromagnetic field is purely electric and radial. Its contribution to the mixed energy--momentum tensor is
\begin{equation}
\begin{aligned}
S^\mu{}_\nu
&=\frac{E^2}{8\pi}\operatorname{diag}(1,1,-1,-1),\\
E(r)&=\frac{q(r)}{r^2},
\end{aligned}
\end{equation}
where $q(r)$ denotes the charge enclosed within radius $r$ and $E(r)$ is the signed radial electric field measured by a static local observer. The field carries a positive energy density $E^2/(8\pi)$, a radial tension of the same magnitude, and a positive transverse pressure. Thus, an electric field contributes to the anisotropy of the total source even when the material fluid itself is isotropic \cite{bekenstein,chargedstability}.

The gravitational field is determined by
$G^\mu{}_\nu=8\pi(T^\mu{}_\nu+S^\mu{}_\nu)$.
For the metric above, the independent components take the form
\begin{align}
8\pi\rho+E^2
&=\frac{1-H}{r^2}-\frac{H'}{r},
\label{eq:rho}\\
8\pi p_r-E^2
&=H\left(\frac{1}{r^2}+\frac{\nu'}{r}\right)
-\frac{1}{r^2},
\label{eq:pr}\\
8\pi p_t+E^2
&=H\left(
\frac{\nu''}{2}
+\frac{\nu'^2}{4}
+\frac{\nu'}{2r}
\right)
+H'\left(
\frac{\nu'}{4}
+\frac{1}{2r}
\right).
\label{eq:pt}
\end{align}
These equations couple the geometry to both the material stresses and the electromagnetic field, as in charged anisotropic stellar constructions \cite{deb2018,maurya}. It is useful to retain this separation explicitly by defining the total energy density and principal pressures as
\begin{equation}
\begin{aligned}
\rho_{\rm tot}&=\rho+\frac{E^2}{8\pi},\\
P_{r,\rm tot}&=p_r-\frac{E^2}{8\pi},\\
P_{t,\rm tot}&=p_t+\frac{E^2}{8\pi}.
\end{aligned}
\end{equation}
In particular, the total pressure anisotropy is
$P_{t,\rm tot}-P_{r,\rm tot}
=\Delta+E^2/(4\pi)$.
Keeping the material and electromagnetic contributions distinct is necessary when imposing an equation of state or evaluating energy conditions for either the material fluid or the complete source.

For the purely radial electric field considered here, the direct Maxwell contribution to the pressure anisotropy is $E^2/(4\pi) \geq 0$. Therefore, a negative value of $\Delta$ is something distinct from that difference in electromagnetic stresses. Nevertheless, the charge can indirectly modify the anisotropy of the material, since the coupled field equations jointly determine both the material profiles and its geometry.

To specify the radial material pressure, we adopt the simplest MIT bag-model relation. In this description, the bag constant $\mathcal{B}$ represents a vacuum energy density that contributes positively to the energy density and negatively to the pressure \cite{farhi,chodos}. For cold, massless, noninteracting quarks, the kinetic contributions satisfy $p_{\rm kin}=\rho_{\rm kin}/3$. Writing
$\rho=\rho_{\rm kin}+\mathcal{B}$ and
$p=p_{\rm kin}-\mathcal{B}$
then gives the familiar linear bag relation. Following its use in anisotropic strange-star models \cite{shee2018,deb2018}, we impose this relation on the radial material sector:
\begin{equation}
p_r=\frac{\rho-4\mathcal{B}}{3},
\qquad \mathcal{B}>0,
\label{eq:eos}
\end{equation}
with $\mathcal{B}$ constant throughout each configuration.

Equation~\eqref{eq:eos} provides a simple description of the radial stiffness while retaining the finite zero-pressure density characteristic of a self-bound model. In particular, at a surface defined by $p_r(R)=0$, it implies $\rho(R)=4\mathcal{B}$. For fixed $\mathcal{B}$, the slope of the radial relation is $dp_r/d\rho=1/3$. Its interpretation as a radial adiabatic sound speed assumes that the same relation governs the corresponding material perturbations. The present prescription neglects finite quark masses and interaction corrections, which enter more detailed descriptions of strange matter \cite{farhi}. Moreover, applying the bag relation to $p_r$ in an anisotropic fluid is a phenomenological choice: it does not supply a microscopic constitutive law for $p_t$, nor does it establish the absolute stability of strange matter. The electromagnetic energy is treated separately and is therefore not included in the variables entering Eq.~\eqref{eq:eos}.

The distinction from an ideal quark gas can also be expressed in terms of the trace of the stress tensor. For massless, non-interacting particles, the kinetic stress tensor has a vanishing trace, even when its momentum distribution is anisotropic; therefore, by adding an isotropic "bag-like" contribution, one obtains $\rho-p_r-2p_t=4\mathcal{B}$. In the present reconstruction, Eq.~\eqref{eq:eos} implies, instead,
\begin{equation}
\rho-p_r-2p_t=4\mathcal{B}-2\Delta.
\label{eq:materialtrace}
\end{equation}

Therefore, wherever $\Delta \neq 0$, the source cannot consist solely of an ideal gas of massless quarks and an isotropic "bag" vacuum. We interpret this as an effective anisotropic fluid with a radial, MIT-bag-type confinement; a microscopic realization would require additional material stresses. The classical Maxwell tensor is traceless and is already accounted for separately; thus, it cannot explain this trace difference, which is of material origin.

The charge distribution is related to the proper charge density $\sigma(r)$ through Maxwell's equations. For a comoving current $J^\mu=\sigma u^\mu$, with
$u^\mu=e^{-\nu/2}\delta^\mu_t$, the radial Maxwell equation becomes
\begin{equation}
q'=\frac{4\pi r^2\sigma}{\sqrt{H}}.
\label{eq:maxwell}
\end{equation}
The factor $H^{-1/2}$ accounts for the proper radial volume element. In the absence of a central point charge, $q(0)=0$, and a finite central charge density together with regular geometry implies $q = \mathcal{O}(r^3)$ and $E = \mathcal{O}(r)$ as $r\rightarrow0$. The electric field consequently vanishes at the centre, ensuring that its contribution to the stress tensor remains regular. A cubic charge prescription is consistent with this behaviour and has been used in charged strange-star models \cite{deb2018}.

Finally, conservation of the total energy--momentum tensor,
$\nabla_\mu(T^\mu{}_\nu+S^\mu{}_\nu)=0$,
gives the charged anisotropic equilibrium equation
\begin{equation}
p_r'
+\frac{\nu'}{2}(\rho+p_r)
-\frac{2(p_t-p_r)}{r}
-\frac{qq'}{4\pi r^4}
=0.
\label{eq:tov}
\end{equation}
This is the hydrostatic balance equation including both anisotropic stresses and the electric force \cite{bowers,bekenstein,deb2018}. Written as an equation for $p_r'$, it displays the gravitational term
$-(\rho+p_r)\nu'/2$, the anisotropic contribution $2\Delta/r$, and the electric contribution $qq'/(4\pi r^4)$.
The anisotropic force acts outward when $\Delta>0$, as does the electric force wherever $qq'>0$. In equilibrium the pressure gradient balances these forces together with gravity. Momentum can be transferred between the fluid and the electromagnetic field through the Lorentz force. We consequently impose conservation on the total tensor, not on each contribution separately.
\section{Compatibility of a Maxwell-only minimal deformation}

We first consider a particular case of minimal geometric deformation: can the extra source be identified entirely with a radial electric field? In the usual radial formulation, the temporal metric potential is retained while the radial geometry is modified by an additional gravitational sector \cite{ovalle}. The physical interpretation of that sector, however, must be consistent with both its stress tensor and the conservation equations. This becomes particularly restrictive when the additional source is required to be purely electromagnetic.

Let $H_0(r)$ describe a regular neutral background with temporal potential $\nu(r)$ and material variables $\rho_0$, $p_{r,0}$ and $p_{t,0}$. We consider the radial deformation
\begin{equation}
H=H_0+g,
\qquad
g=\alpha f,
\end{equation}
where $\alpha$ controls the deformation amplitude.

\noindent\textbf{Proposition (deformation restricted to a Maxwell field).} Let the neutral reference state and the deformed configuration be regular static solutions defined on the same connected interior domain $0\leq r\leq R$, with $H_0(0)=H(0)=1$ and $0<e^{\nu(0)}<\infty$. Assume that the temporal potential $\nu$ and all material components $(\rho,p_r,p_t)$ remain invariant, and that the additional source consists exclusively of a classical radial Maxwell field. Then, the deformation and the electric field vanish identically: $g\equiv0$ and $E\equiv0$.

\noindent\textit{Proof.} Under these assumptions, the difference between the deformed and background Einstein equations must be accounted for entirely by the Maxwell tensor.

Subtracting the background equations from Eqs.~\eqref{eq:rho} and \eqref{eq:pr} gives
\begin{equation}
\begin{aligned}
E^2&=-\frac{g'}{r}-\frac{g}{r^2},\\
E^2&=-g\left(\frac{\nu'}{r}+\frac{1}{r^2}\right).
\end{aligned}
\end{equation}
The two expressions must describe the same electric field. Equating them yields
\begin{equation}
g'=\nu'g,
\qquad
g=C e^\nu,
\label{eq:nogo}
\end{equation}
where $C$ is an integration constant. This condition follows from the equality of the temporal and radial mixed components of the Maxwell tensor and is already sufficient to constrain the proposed deformation.

For a regular centre, both the original and deformed radial potentials must satisfy $H_0(0)=H(0)=1$. Consequently, $g(0)=0$. Since the temporal potential is finite and nonzero at the centre, Eq.~\eqref{eq:nogo} then requires $C=0$. Thus, within the assumptions stated above, the only regular solution is
\[
g(r)=0,
\qquad
E(r)=0.
\]
The obstruction arises from requiring the geometry to accommodate an electric field while keeping both the temporal potential and the neutral material distribution fixed.

The same conclusion can be understood from the conservation equations. For a static anisotropic fluid, the radial material conservation equation depends on $\nu$, $\rho$, $p_r$ and $p_t$, but not explicitly on $H$. Keeping these quantities equal to those of the neutral background therefore preserves the background material conservation equation. Conservation of the total source would then require the Maxwell sector to be separately conserved. Its radial conservation equation reduces to
\begin{equation}
(E^2)'+\frac{4E^2}{r}=0,
\end{equation}
so that
\[
r^4E^2=q^2=\mathrm{constant}.
\]
A nonzero constant corresponds to a Coulomb field that is singular at the centre. Regularity therefore again selects the vanishing-charge solution. Physically, a distributed charged fluid generally exchanges momentum with the electric field through the Lorentz force, as reflected in the charged equilibrium equation \cite{bekenstein}. Requiring the original neutral material balance to remain unchanged removes this possibility.

This obstruction is consistent with the independent conservation constraint inherent to minimal radial decoupling. Extended geometric deformation relaxes this constraint by allowing the temporal metric potential to vary; its recovery of the Reissner-Nordström geometry provides an explicit electromagnetic example \cite{ovalle2019decoupling}. The proposition pertains to the fixed neutral material source and the regular center specified earlier, and not to the existence of charged solutions within the broader framework of decoupling.

This result applies specifically to a regular interior containing the centre and to a deformation whose entire additional source is Maxwellian. It does not apply to an electrovacuum exterior, which need not contain a regular centre. Nor does it exclude constructions that begin with a charged background and introduce a separate anisotropic sector \cite{morales2018charged,albalahi}. Allowing the temporal potential to change, or determining the material distribution together with the electric field, also lies outside the assumptions leading to Eq.~\eqref{eq:nogo}. These distinctions motivate the coupled reconstruction developed below.

\section{Exact reconstruction of a regular charged baseline}

We now determine the material variables and radial geometry directly from the coupled Einstein--Maxwell system. As a geometric input, we retain the temporal dependence associated with the Tolman-IV solution \cite{tolman},
\begin{equation}
e^\nu=b^2\left(1+\frac{r^2}{a^2}\right),
\qquad
a>0,
\quad
b>0.
\end{equation}
Here $a$ sets an interior length scale, while $b$ fixes the normalization of the time coordinate and will be determined by surface matching. Only the temporal potential is prescribed: the radial potential is reconstructed from the field equations and the radial bag relation. The resulting configuration is therefore not assumed to retain the isotropy or the full metric of the original Tolman-IV solution.

For the electric charge, we adopt
\begin{equation}
\begin{aligned}
q(r)&=\frac{\sqrt{\varepsilon}\,r^3}{a^2},\\
E^2(r)&=\frac{\varepsilon r^2}{a^4},
\qquad
\varepsilon\geq0.
\end{aligned}
\label{eq:charge}
\end{equation}
A cubic charge profile has previously been used in charged strange-star models \cite{deb2018}. Its usefulness here is that it satisfies the central behaviour required by a finite proper charge density, while introducing a single dimensionless parameter controlling the field strength. The charge parameter $\varepsilon$ should be distinguished from the deformation coupling $\alpha$ used in the preceding section. We choose the positive sign of $q$; reversing the sign of the entire charge distribution leaves the metric and material stresses unchanged.

Equation~\eqref{eq:charge} is a prescribed equilibrium charge distribution rather than a microscopic model for charge separation. In particular, it does not imply a spatially constant proper charge density. Substitution into Eq.~\eqref{eq:maxwell} gives
\begin{equation}
\sigma(r)=
\frac{3\sqrt{\varepsilon}\sqrt{H(r)}}{4\pi a^2}.
\end{equation}
Thus, the proper charge density depends on the reconstructed geometry and remains finite at the centre whenever the metric is regular.

The radial bag relation now provides a direct equation for $H$. Combining Eqs.~\eqref{eq:rho}, \eqref{eq:pr} and \eqref{eq:eos}, we obtain
\begin{equation}
H'
+\left(\frac{4}{r}+3\nu'\right)H
=
\frac{4}{r}
-32\pi\mathcal{B}r
-4rE^2.
\label{eq:ode}
\end{equation}
For the prescribed temporal potential, this is a first-order linear equation with integrating factor proportional to
$r^4(a^2+r^2)^3$. Its homogeneous contribution is proportional to
$[r^4(a^2+r^2)^3]^{-1}$ and diverges at the centre. Removing this contribution selects the unique regular radial solution for the chosen values of $a$, $\mathcal{B}$ and $\varepsilon$. The remaining Einstein equation then determines the tangential material pressure.

To separate the shape of the solution from its dimensional scale, we introduce
\[
x=\frac{r}{a},
\qquad
t=x^2,
\qquad
\beta=8\pi\mathcal{B}a^2.
\]
The regular solution of Eq.~\eqref{eq:ode} can then be written as
\begin{align}
H(t)&=\frac{N(t)}{D(t)},
\qquad
D(t)=30(1+t)^3,
\label{eq:h}\\
N(t)&=N_0(t)+\beta N_B(t)+\varepsilon N_E(t),\\
N_0(t)&=30+60t+45t^2+12t^3,\\
N_B(t)&=-20t-45t^2-36t^3-10t^4,\\
N_E(t)&=-15t^2-36t^3-30t^4-\frac{60}{7}t^5.
\end{align}
This representation separates the charge-independent terms from the contributions proportional to the bag parameter and the electric-field amplitude. Since $D(t)>0$ for $t\geq0$, the sign of the radial metric potential is controlled entirely by $N(t)$, a useful feature for the subsequent admissibility analysis.

In the neutral limit, $\varepsilon=0$, the electromagnetic contribution vanishes and the solution describes an anisotropic configuration governed by the radial bag relation and the prescribed temporal potential. This limit does not generally recover the complete isotropic Tolman-IV solution, since the radial geometry is determined by Eq.~\eqref{eq:ode}. For nonzero charge, the same coupled equation determines how the radial metric and material variables change in response to the electric field. The reconstruction therefore allows the material source to adjust together with the geometry, consistently with the coupled treatment used in charged bag-model configurations \cite{deb2018,maurya,goswami}.

For a compact expression of the material variables, define
\[
K(t)=\frac{N(t)-D(t)}{t},
\]
where the value at $t=0$ is understood through its polynomial continuation. This continuation exists because $N(0)=D(0)$. Introducing dimensionless densities and pressures, the field equations become
\begin{align}
\widehat{\rho}
&\equiv 8\pi a^2\rho
=-\frac{K}{D}-2H_t-\varepsilon t,
\label{eq:rhodim}\\
\widehat{p}_r
&\equiv 8\pi a^2p_r
=\frac{K}{D}+\frac{2H}{1+t}+\varepsilon t,
\label{eq:prdim}\\
\widehat{p}_t
&\equiv 8\pi a^2p_t
=\frac{H(2+t)}{(1+t)^2}
+\frac{(1+2t)H_t}{1+t}
-\varepsilon t,
\label{eq:ptdim}
\end{align}
where $H_t=dH/dt$. All three quantities refer to the material sector; the electromagnetic stresses remain separate.

The reconstructed metric satisfies the differential identity
\begin{equation}
\begin{aligned}
2t(1+t)H_t+(4+10t)H
={}&(4-4\beta t-4\varepsilon t^2)(1+t).
\end{aligned}
\end{equation}
Together with Eqs.~\eqref{eq:rhodim} and \eqref{eq:prdim}, this gives
$\widehat{p}_r=(\widehat{\rho}-4\beta)/3$ identically. The radial equation of state is therefore satisfied throughout the interior, rather than only at selected radii or at the surface.

The central behaviour follows by expanding around $x=0$:
\begin{align}
H&=1-\left(1+\frac{2\beta}{3}\right)x^2 + \mathcal{O}(x^4),\\
\widehat{\rho}_c&=3+2\beta,\\
\widehat{p}_{r,c}
&=\widehat{p}_{t,c}
=1-\frac{2\beta}{3}.
\end{align}
At fixed $\beta$, the electric field does not contribute directly to these central values because $E^2$ vanishes quadratically with radius. Once the surface conditions are imposed, however, $\beta$ depends on $\varepsilon$ and the central values may change with charge. With $\mathcal{B}>0$, the condition of positive central pressure reduces to $0<\beta<3/2$. The two pressures coincide at the centre, as required by spherical symmetry and regularity. This equality does not determine the anisotropy at finite radius.

\section{Boundary matching and physical admissibility}
\label{sec:certificate}

To interpret the interior as a finite stellar configuration, we must specify its boundary and match it to an exterior solution. We consider a surface at $r=R$ separating the charged material distribution from a Reissner--Nordstr\"om exterior with mass $M$ and charge $Q$. In the absence of a thin shell, the induced metric and extrinsic curvature must be continuous across this surface \cite{israel1966}. We also assume that no charge is concentrated on the boundary, so that $q(R)=Q$.

With a common normalization of the time coordinate, these conditions give
\begin{equation}
\begin{gathered}
H(R)=e^{\nu(R)}=F_R,\\
F_R=1-\frac{2M}{R}+\frac{Q^2}{R^2},
\qquad
p_r(R)=0.
\end{gathered}
\end{equation}
The vanishing pressure in this expression is the material radial pressure. The electric field extends beyond the star, and its radial stress therefore remains present on both sides of the boundary. Continuity of the charge ensures continuity of this electromagnetic contribution, leaving $p_r(R)=0$ as the material matching condition \cite{bekenstein,deb2018}. The tangential material pressure need not vanish at the surface: a finite value of $p_t(R)$ is compatible with the absence of a thin shell.

The temporal normalization follows immediately,
\begin{equation}
b^2=\frac{F_R}{1+R^2/a^2}.
\end{equation}
To express the remaining conditions without fixing the stellar radius relative to the interior length scale, we introduce
\[
y=\left(\frac{R}{a}\right)^2>0.
\]
The radial pressure condition, evaluated using Eq.~\eqref{eq:prdim}, then yields
\begin{equation}
H(R)=
\frac{(1-\varepsilon y^2)(1+y)}{1+3y}.
\end{equation}
Consequently, a positive surface metric requires
$\varepsilon y^2<1$. Equating this boundary value to the reconstructed potential in Eq.~\eqref{eq:h} also fixes the dimensionless bag parameter. Thus, the charge amplitude, surface-to-scale ratio and bag parameter are linked by the matching conditions and cannot all be chosen independently.

For compactness of notation, define
\begin{align}
A(y)&=10y^3+36y^2+45y+20,\\
C(y)&=10y^5+50y^4+98y^3+91y^2+35y,\\
P(y)&=14y^3+63y^2+105y+70.
\end{align}
The resulting parameter relation is
\begin{equation}
\beta(y,\varepsilon)=
\frac{3[P(y)+\varepsilon C(y)]}
{7(3y+1)A(y)}.
\label{eq:generalbeta}
\end{equation}
All three polynomials are positive for $y>0$, so the matched bag parameter is positive whenever $\varepsilon\geq0$. This does not yet guarantee an acceptable interior, since conditions such as positive central pressure impose additional restrictions. If the dimensional bag constant $\mathcal{B}$ is prescribed, Eq.~\eqref{eq:generalbeta} determines the length scale through
$a^2=\beta/(8\pi\mathcal{B})$, after which $R=a\sqrt{y}$.

The connection with the exterior mass is conveniently expressed through the charged mass function,
\begin{equation}
\begin{aligned}
m(r)&=\frac{r}{2}
\left[1-H(r)+\frac{q(r)^2}{r^2}\right],\\
m'&=4\pi r^2\rho+\frac{qq'}{r}.
\end{aligned}
\end{equation}
This definition gives $m(R)=M$ and a constant mass parameter in the electrovacuum exterior. Its derivative displays the material and charge-dependent contributions to the mass balance \cite{bekenstein}. Combining the surface metric with the charge prescription gives
\begin{equation}
\begin{aligned}
\frac{M}{R}
&=\frac{y+\varepsilon y^2(1+2y)}{1+3y},\\
\frac{Q}{R}
&=\sqrt{\varepsilon}\,y.
\end{aligned}
\end{equation}
These relations allow the equilibrium family to be explored in terms of $y$ and $\varepsilon$ while satisfying the boundary conditions exactly. For $\varepsilon=0$, writing $u=M/R$ gives $e^{\nu}=1-3u+u(r/R)^2$, as in Ref.~\cite{santana2026radial}. Since the radial equation and central regularity condition also coincide, both constructions yield the same neutral geometry.

The tangential pressure provides a further restriction that is not apparent from metric matching alone. Introducing
\begin{equation}
\begin{aligned}
J(y)={}&540y^5+2662y^4+5200y^3\\
&+4977y^2+2261y+350,
\end{aligned}
\end{equation}
the material tangential pressure at the surface becomes
\begin{equation}
\widehat{p}_t(R)=
\frac{
y[21(2y^3+6y^2+3y-5)-\varepsilon J(y)]
}{
7(1+y)(3y+1)A(y)
}.
\label{eq:generalpt}
\end{equation}
Since the denominator and $J(y)$ are positive for $y>0$, nonnegative tangential surface pressure requires
\begin{equation}
0\leq\varepsilon\leq
\frac{21(2y^3+6y^2+3y-5)}{J(y)}.
\label{eq:generalbound}
\end{equation}
This inequality expresses a requirement on the material configuration, rather than an additional junction condition.

The numerator in Eq.~\eqref{eq:generalbound} also restricts the allowed surface-to-scale ratio. The polynomial
$2y^3+6y^2+3y-5$ is strictly increasing for $y>0$ and has a unique positive root,
$y_0\simeq0.6474266573$. Therefore, no configuration with $0<y<y_0$ satisfies the adopted nonnegative tangential surface-pressure condition, even at zero charge. At $y=y_0$, that condition permits only the neutral case. For larger $y$, it supplies an upper charge limit that must be considered together with $\varepsilon y^2<1$ and the requirements throughout the interior. Section~\ref{sec:survey} establishes that the surface bound is also sufficient for static admissibility.

A particularly transparent case is obtained by setting $y=1$, or $R=a$. The matching relations then reduce to
\begin{equation}
\begin{gathered}
\beta=\frac{9}{37}+\frac{71}{259}\varepsilon,
\qquad
b^2=\frac{1-\varepsilon}{4},\\
\frac{Q}{R}=\sqrt{\varepsilon},
\qquad
\frac{M}{R}=\frac{1}{4}+\frac{3}{4}\varepsilon.
\end{gathered}
\label{eq:matchone}
\end{equation}
On this branch, increasing $\varepsilon$ changes the compactness, the charge-to-radius ratio and the matched bag parameter together. Substituting Eq.~\eqref{eq:matchone} into the tangential pressure gives
\begin{equation}
\widehat{p}_t(R)
=\frac{3}{148}-\frac{2665}{1036}\varepsilon,
\end{equation}
and hence
\begin{equation}
0\leq\varepsilon\leq
\varepsilon_*\equiv\frac{21}{2665}
\simeq0.007880.
\label{eq:bound}
\end{equation}
The limiting configuration has $p_t(R)=0$, whereas the tangential surface pressure is positive below this limit. Thus, the charge restriction arises before the surface metric loses positivity and reflects the response of the material stresses to the electric field.

To determine whether this surface restriction is also sufficient for an acceptable interior, we examine the full interval $0\leq t\leq1$. Here, static admissibility refers to a regular geometry with positive metric potentials, nonnegative material density and pressures, nonincreasing material profiles, and satisfaction of the weak, dominant and strong energy conditions. These requirements constrain the equilibrium configuration; they do not establish stability against time-dependent perturbations or supply a microscopic description of the anisotropic matter \cite{horvat,anisostability,chargedstability}. The derivative ratios are also examined, with their interpretation restricted to the equilibrium profiles.

For the matched $R=a$ branch, Eq.~\eqref{eq:bound} is both necessary and sufficient for nonnegative tangential material pressure throughout the star. Moreover, every configuration in this interval satisfies the static conditions just stated. This can be established analytically because, after substituting Eq.~\eqref{eq:matchone}, the relevant rational expressions have a simple dependence on the charge parameter.

Specifically, $H$, $\widehat{\rho}$, $\widehat{p}_r$,
$\widehat{p}_t$, the signed derivatives listed in
Table~\ref{tab:certificate}, and the dominant-energy margins are affine functions of $\varepsilon$ at fixed $t$. If $F$ denotes any one of these quantities, then
\begin{equation}
\begin{aligned}
F(t,\varepsilon)
={}&(1-\eta)F(t,0)
+\eta F(t,\varepsilon_*),\\
\eta={}&\frac{\varepsilon}{\varepsilon_*}\in[0,1].
\end{aligned}
\end{equation}
It is therefore enough to establish the required signs at the two charge endpoints. Positivity at these endpoints extends to the entire interval by convexity.

At each endpoint, the expressions can be written as
$P(t)/[d(1+t)^k]$, with a positive denominator and a polynomial numerator. Here $P(t)$ denotes a generic numerator polynomial, distinct from the auxiliary polynomial $P(y)$ introduced in the matching calculation. To determine its sign over the complete radial interval, write
\[
P(t)=\sum_{j=0}^{n}a_jt^j
\]
in the Bernstein basis:
\begin{equation}
\begin{aligned}
P(t)&=\sum_{i=0}^{n}c_i
\binom{n}{i}t^i(1-t)^{n-i},\\
c_i&=\sum_{j=0}^{i}
a_j\frac{\binom{i}{j}}{\binom{n}{j}}.
\end{aligned}
\end{equation}
Each basis function is nonnegative on $[0,1]$. Nonnegative Bernstein coefficients therefore provide an exact sufficient condition for the numerator to remain nonnegative throughout the star, without relying on a finite sampling of the radial interval.

Table~\ref{tab:certificate} reports the minimum Bernstein coefficient for each endpoint numerator. The coefficients are evaluated with exact rational arithmetic. Their signs establish positivity of the radial metric potential and density, nonnegativity of both pressures, and positivity of the indicated derivative and energy-condition margins. Appendix~\ref{app:certificates} specifies the numerator polynomials, their degrees, and the exact construction of the coefficients used to reproduce this table and its extensions to the other five surface-to-scale ratios.

\begin{table}[t]
\centering
\caption{Exact positivity certificates for the matched $R=a$
branch. Each endpoint expression is written as
$P(t)/[d(1+t)^k]$, with $d=37$ at $\varepsilon=0$ and
$d=2665$ at $\varepsilon=\varepsilon_*$. The entries give
the minimum Bernstein coefficient of the corresponding
numerator. Matter variables are dimensionless; hats are
omitted for brevity. A subscript $t$ denotes differentiation
with respect to $t$.}
\label{tab:certificate}
\begin{tabular}{lrrr}
\toprule
Quantity & $k$ & $\varepsilon=0$ & $\varepsilon=\varepsilon_*$\\
\midrule
$H$                       & 3 & 37  & 2665\\
$\rho$                    & 4 & 129 & 9303\\
$p_r$                     & 4 & 0   & 0\\
$p_t$                     & 5 & 24  & 0\\
$-\rho_t$                 & 5 & 300 & 21591\\
$-(p_r)_t$                & 5 & 100 & 7197\\
$-(p_t)_t$                & 6 & 120 & 8691\\
$\rho-p_r$                & 4 & 98  & 7074\\
$\rho-p_t$                & 5 & 98  & 7074\\
$(p_t)_t-\rho_t$          & 6 & 180 & 12900\\
\bottomrule
\end{tabular}
\end{table}

The strictly positive derivative margins show that the density and both pressures decrease with $t$, and hence with $r$ for $r>0$. Their radial derivatives vanish at the centre, consistently with the regular central expansions. Since $p_r(R)=0$, strict decrease implies $p_r>0$ in the interior. Similarly, $p_t$ remains positive inside the star even for the limiting configuration $\varepsilon=\varepsilon_*$, where it vanishes only at the surface. The central expansions already establish $H(0)=1$ and $p_r(0)=p_t(0)$, while Eq.~\eqref{eq:matchone} guarantees a positive temporal normalization throughout the permitted charge interval.

The energy conditions follow directly from the same inequalities. For the material source, the weak condition requires
$\rho\geq0$ and $\rho+p_i\geq0$; the dominant condition requires
$\rho\geq|p_i|$, with $i=r,t$; and the strong condition requires
$\rho+p_r+2p_t\geq0$ together with $\rho+p_i\geq0$.
The nonnegative density and pressures establish the weak and strong conditions, while the positive margins
$\rho-p_r$ and $\rho-p_t$ establish the dominant condition.
These statements concern the material variables defined in the field equations.

The derivative bounds can likewise be obtained without numerical differentiation. The radial bag relation gives
\[
\frac{dp_r}{d\rho}=\frac{1}{3},
\]
whereas the inequalities
$\rho_t<0$, $(p_t)_t<0$ and $(p_t)_t-\rho_t>0$ imply
\begin{equation}
0<
\frac{(p_t)_t}{\rho_t}
<1.
\end{equation}
This ratio describes how the tangential pressure varies with density along the equilibrium profile. Identifying it with a tangential characteristic speed would require a constitutive prescription for perturbations, which is not supplied by the static reconstruction \cite{horvat,pretel}. The certificate therefore establishes equilibrium inequalities without invoking a pulsation operator.

The permitted charge interval corresponds to
\begin{equation}
\begin{gathered}
\frac{1}{4}\leq\frac{M}{R}
\leq\frac{682}{2665}
\simeq0.255910,\\
0\leq\frac{Q}{R}
\leq\sqrt{\frac{21}{2665}}
\simeq0.088769.
\end{gathered}
\end{equation}
These bounds characterise the matched $R=a$ branch under the stated material admissibility requirements. The neutral member is already anisotropic, and the charged configurations need not represent equilibria of a common microscopic material model. The result nevertheless provides a continuous family for which the boundary conditions and the specified interior inequalities are established exactly, furnishing a controlled background for subsequent numerical and perturbative investigations.

The analytical bounds obtained above can be illustrated by considering a representative configuration within the admissible interval. We choose $\varepsilon=10^{-3}$ and $R=a$, for which the matching conditions give $\beta=63071/259000$. This example allows the radial behaviour of the solution to be examined and provides a numerical check of the analytical expressions. We use these parameters solely to illustrate the equilibrium solution, without fitting an observed star.

For the numerical evaluation we used a uniform grid of 10,001 points in $x=r/a\in[0,1]$. At the central point we evaluated the polynomial continuations. We differentiated the profiles with respect to $t=x^2$ by the complex-step method, taking a step of $10^{-25}$. The rational expressions have no poles in the interval, so this approach permits differentiation without the subtraction of nearly equal real values encountered in ordinary finite differences. Evaluating the pressure--density derivative ratios through derivatives with respect to $t$ also gives their regular central limits, where the corresponding radial derivatives vanish simultaneously.

\begin{table}[t]
\centering
\caption{Representative equilibrium configuration with
$\varepsilon=10^{-3}$ and $R=a$. Hatted densities and
pressures are normalized by $8\pi a^2$. All entries are
model predictions for the prescribed parameters.}
\label{tab:representative}
\begin{tabular}{lr}
\toprule
Quantity & Value\\
\midrule
$\beta$ & 0.243517375\\
$M/R$ & 0.250750\\
$Q/R$ & 0.031622777\\
$H(R)$ & 0.499500\\
$b^2$ & 0.249750\\
$\widehat{\rho}(0)$ & 3.487034749\\
$\widehat{p}_r(0)=\widehat{p}_t(0)$ & 0.837655084\\
$\widehat{\rho}(R)$ & 0.974069498\\
$\widehat{p}_r(R)$ & 0\\
$\widehat{p}_t(R)$ & 0.017697876\\
$z_s$ & 0.414921200\\
\bottomrule
\end{tabular}
\end{table}

The values in Table~\ref{tab:representative} are consistent with the regularity and matching conditions. The radial metric potential remains positive, decreasing from $H(0)=1$ to $H(R)=0.4995$, while the material density and both pressures decrease outward. The radial pressure vanishes at the surface, with an absolute numerical boundary error below $10^{-16}$ in the normalized expression. The tangential pressure remains positive there, as expected for a charge parameter below $\varepsilon_*$. The finite surface density satisfies $\widehat{\rho}(R)=4\beta$, in agreement with the radial bag relation.

The material anisotropy has a less uniform behaviour. Although it vanishes at the centre, $\widehat{\Delta}=\widehat{p}_t-\widehat{p}_r$ takes both signs within the configuration. Its sampled minimum is approximately $-0.03665445$, while its maximum is approximately $0.01769788$. Through the term $2\Delta/r$ in the equilibrium equation, the material anisotropy therefore contributes inward in some regions and outward in others. This behaviour illustrates why the support of the configuration must be understood from the combined pressure, gravitational, electric and anisotropic contributions, rather than from the sign of any one term alone \cite{bowers,bekenstein}.

The numerical energy-condition checks agree with the continuous certificate established above. The smallest sampled value among the normalized material dominant-energy margins,
$\widehat{\rho}-|\widehat{p}_r|$ and
$\widehat{\rho}-|\widehat{p}_t|$, is approximately
$0.95637162$. The weak and strong conditions are also satisfied. These statements extend to the total source in this example: the Maxwell contribution satisfies the same standard energy conditions, and adding it to the material tensor preserves them. In particular, the strong condition involves
$\rho+p_r+2p_t\geq0$, together with
$\rho+p_r\geq0$ and $\rho+p_t\geq0$; the expression
$\rho-p_r-2p_t$ is a different combination and is not used here as the strong-energy criterion.

The pressure--density derivatives provide further information about the equilibrium profiles. The radial relation fixes $dp_r/d\rho=1/3$ exactly, whereas the sampled tangential ratio
$(p_t)_t/\rho_t$ ranges from approximately $0.27342163$ to $0.40032070$. These values agree with the analytical bounds, but their physical interpretation requires care. The tangential ratio describes variation along the reconstructed background and becomes a characteristic propagation speed only after an appropriate constitutive law for perturbations has been specified \cite{horvat,pretel}.

A related diagnostic is the radial adiabatic-index expression
\begin{equation}
\Gamma_r=
\frac{\rho+p_r}{p_r}\frac{dp_r}{d\rho}.
\end{equation}
For the present example, its sampled minimum is approximately $1.72095091$. It diverges as the surface is approached because $p_r$ tends to zero while the material density remains finite. This behaviour follows from the self-bound form of the adopted radial equation of state and does not by itself establish relativistic dynamical stability. In addition, the difference
$(p_t)_t/\rho_t-1/3$ changes sign across the sampled profile. A classification based on a fixed sign throughout the star is therefore unavailable for these derivative ratios, and a perturbative treatment would be needed to assess the response of the configuration \cite{andrade_cracking2022,dev,anisostability}.

The temporal potential also determines the gravitational redshift of a photon emitted by a static source and received by an observer at infinity,
\begin{equation}
z(r)=e^{-\nu(r)/2}-1,
\end{equation}
where the time coordinate is normalized by the exterior matching. Since $e^\nu=b^2(1+r^2/a^2)$ increases outward, the redshift decreases from the centre to the surface. For the chosen configuration,
$z_s=F_R^{-1/2}-1\simeq0.414921200$.
This is a prediction of the matched geometry and is not used here to identify an observed object as a strange star.

The numerical implementation was also checked directly against the governing equations. The maximum absolute residuals of the normalized bag relation and metric differential equation are below $4\times10^{-15}$. The normalized charged-equilibrium residual is below $5\times10^{-12}$, with the largest sensitivity associated with cancellation in the anisotropy term near the centre. The small residuals measure the accuracy of the floating-point implementation, independently of the polynomial identity for the radial potential. It is the analytical certificate, not the number of grid points, that guarantees the inequalities at every radius.

An example outside the allowed charge interval shows that the field equations can still be satisfied even when a material admissibility condition fails. At $R=a$ and $\varepsilon=10^{-2}$, the matched expressions remain well defined, but they give
$\widehat{p}_t(R)\simeq-0.00545367$.
This configuration therefore fails the adopted nonnegative tangential-pressure condition. The example shows that a regular metric and an exactly satisfied radial equation of state are not sufficient to ensure all of the required properties of the material source.

As an independent check of the closed-form metric, Eq.~\eqref{eq:ode} was integrated numerically after expressing it in terms of $t$. The integration used the DOP853 method and began at $t=10^{-6}$, with initial data obtained from the fourth-order central series. For
$\varepsilon=0$, $10^{-3}$, $21/2665$ and $10^{-2}$, comparison with the analytical solution at 2,001 points showed a maximum metric discrepancy below $2.6\times10^{-8}$ at a relative tolerance $10^{-7}$, decreasing to below $4.5\times10^{-12}$ at a relative tolerance $10^{-12}$. In both cases, the absolute tolerance was set to one hundredth of the relative tolerance. The agreement includes the deliberately excluded configuration at $\varepsilon=10^{-2}$, confirming that numerical accuracy and physical admissibility are separate issues.

A dimensional interpretation can be assigned once a physical value of the bag energy density is specified. Restoring units gives
\begin{equation}
\begin{aligned}
\mathcal{B}_{\rm geom}
&=\frac{G\mathcal{B}_{\rm SI}}{c^4},\\
a&=\left(
\frac{\beta}{8\pi\mathcal{B}_{\rm geom}}
\right)^{1/2}.
\end{aligned}
\end{equation}
Here $\mathcal{B}_{\rm SI}$ is an energy density. For a fixed dimensionless configuration, the stellar radius and physical mass therefore scale as $\mathcal{B}_{\rm SI}^{-1/2}$, with the mass in ordinary units obtained from
$M_{\rm SI}=c^2M_{\rm geom}/G$.

For illustration, choosing
$\mathcal{B}=60\,\mathrm{MeV\,fm^{-3}}$
for the configuration with $\varepsilon=10^{-3}$ and $R=a$ gives
$R\simeq11.045\,\mathrm{km}$ and
$M\simeq1.876\,M_\odot$.
This assignment provides a dimensional example of the equilibrium solution, rather than a determination of the bag constant from observations.

The corresponding electric charge is obtained from
\begin{equation}
Q_{\rm SI}
=
\frac{c^2\sqrt{4\pi\epsilon_0}}{\sqrt{G}}\,
Q_{\rm geom},
\end{equation}
where $\epsilon_0$ is the vacuum permittivity, distinct from the dimensionless charge parameter $\varepsilon$. The illustrative scaling gives
$Q\simeq4.053\times10^{19}\,\mathrm{C}$
and a surface electric field of approximately
$2.986\times10^{21}\,\mathrm{V\,m^{-1}}$.
Thus, a small dimensionless charge parameter need not correspond to a weak electric field in physical units. Comparable charge and electric-field scales arise in charged-fluid equilibrium calculations \cite{chargedstability}. Their interpretation requires additional physics governing charge separation, screening and retention, none of which is fixed by the prescribed charge profile.

\begin{revision}
The surface field obtained in this example is unusually large. To put it into perspective, we compare it with the critical field of quantum electrodynamics,
\begin{equation}
E_{\rm cr}=\frac{m_e^2c^3}{e_{\rm el}\hbar}
\simeq1.32\times10^{18}\,\mathrm{V\,m^{-1}},
\label{eq:criticalfield}
\end{equation}
where $m_e$ and $e_{\rm el}>0$ are the electron mass and elementary charge \cite{schwinger1951}. Pair creation in a constant electric field is exponentially suppressed for $E\ll E_{\rm cr}$; there is no sharp threshold at $E=E_{\rm cr}$. In our example, however, $E(R)/E_{\rm cr}\simeq2.3\times10^3$. At such fields, the classical equilibrium calculation alone cannot settle whether the assumed charge can be maintained.

Outside the star, the Reissner--Nordstr\"om solution gives the following field magnitude for a local static observer:
\begin{equation}
\begin{aligned}
E_{\rm ext}(r)&=\frac{|Q_{\rm SI}|}{4\pi\epsilon_0r^2}\\
&=E(R)\left(\frac{R}{r}\right)^2,
\qquad r\geq R.
\end{aligned}
\label{eq:exteriorfieldSI}
\end{equation}
Both radii are expressed in SI length units. For $E(R)>E_{\rm cr}$, the classical field falls to $E_{\rm cr}$ at
\begin{equation}
r_{\rm cr}=R\sqrt{\frac{E(R)}{E_{\rm cr}}}.
\label{eq:criticalradius}
\end{equation}
Using $R\simeq11.045\,\mathrm{km}$ and
$E(R)\simeq2.986\times10^{21}\,\mathrm{V\,m^{-1}}$
gives $r_{\rm cr}\simeq5.25\times10^2\,\mathrm{km}\simeq47.6R$.
The field therefore remains above the critical scale well beyond the stellar surface, throughout $R\leq r<r_{\rm cr}$. We stress that this estimate follows from the assumed electrovacuum solution. It tells us neither where a plasma would end nor how rapidly the star would discharge.

The situation is quite different from that of the electrosphere of a bare strange star. Usov et al.~\cite{usov2005electrospheres} describe a surface electron layer with a characteristic thickness of order $10^2$--$10^3\,\mathrm{fm}$, whose field and particle distributions are determined together. The occupation of electron states and the thermal creation of vacancies regulate pair production in that layer. In our model, by contrast, the region extending to $r_{\rm cr}$ in Eq.~\eqref{eq:criticalradius} is treated as electrovacuum. Even if the surface fields are comparable, there is no reason to assume that the same suppression mechanisms operate over this much larger region.

In SI units, $E_{\rm SI}=c^2E_{\rm geom}/\sqrt{4\pi\epsilon_0G}$. Using $E_{\rm geom}^2(R)=\varepsilon y/a^2$ and the scale fixed by the bag parameter, we obtain
\begin{equation}
E_{\rm SI}^2(R)=\frac{2\mathcal{B}_{\rm SI}}{\epsilon_0}
\frac{\varepsilon y}{\beta(y,\varepsilon)}.
\label{eq:surfacefieldbag}
\end{equation}
The field is maximal at $R$. Imposing $E_{\rm SI}(R)\leq E_{\rm cr}$ gives
\begin{equation}
\frac{\varepsilon y}{\beta(y,\varepsilon)}\leq \kappa,
\qquad
\kappa=\frac{\epsilon_0 E_{\rm cr}^2}{2\mathcal{B}_{\rm SI}}.
\label{eq:schwingercriterion}
\end{equation}
From Eq.~\eqref{eq:generalbeta}, $\beta=\beta_0(y)+\varepsilon\beta_1(y)$, with $\beta_0=\beta(y,0)$ and $\beta_1=\partial\beta/\partial\varepsilon$. Hence
\begin{equation}
\varepsilon\leq\frac{\kappa\beta_0(y)}{y-\kappa\beta_1(y)},
\quad y>\kappa\beta_1(y).
\label{eq:schwingerbound}
\end{equation}
For $y\leq\kappa\beta_1(y)$, this condition adds no charge bound. At $y=1$ and $\mathcal{B}=60\,\mathrm{MeV\,fm^{-3}}$, $\kappa\simeq8.0\times10^{-10}$ gives $\varepsilon\lesssim1.95\times10^{-10}$, well below the classical limit $21/2665\simeq7.88\times10^{-3}$. Pair production can occur even below the Schwinger scale, as Page \cite{page2006dyadospheres} showed for collapsing charged cores. Our static model does not describe this discharge.

Whether a star could retain this charge is an open question. To answer it, one would need to calculate the current produced by the pairs and account for charge transport, plasma screening and the resulting change in the electric field. None of these processes enters our static solution. If an atmosphere or an electrosphere were included, the exterior solution and the matching conditions would also have to be reconsidered. The certified intervals should thus be understood as a result of the classical Einstein--Maxwell calculation: they establish regularity and material admissibility, but not quantum-electrodynamic consistency or charge retention over stellar timescales.
\end{revision}

Taken together, the analytical bounds, equation residuals and independent integration establish a consistent equilibrium example and clarify the limits of its interpretation. We now establish the general admissibility domain and examine six representative families.

\section{Physical acceptability across compactness and charge}
\label{sec:survey}

The preceding construction establishes an admissible charged branch at $R=a$. To examine how its properties change when the surface-to-scale ratio is varied, we now consider the two-parameter family described by $y=(R/a)^2$ and $\varepsilon$. At every point, the bag parameter, temporal normalization and material profiles are recalculated from the matching conditions. The comparison therefore concerns complete equilibrium configurations, each satisfying the same radial bag prescription and charge law.

The exterior quantities are related to the interior parameters through
\begin{equation}
\begin{aligned}
u\equiv\frac{M}{R}
&=\frac{y+\varepsilon y^2(1+2y)}{1+3y},\\
\frac{Q}{R}&=\sqrt{\varepsilon}\,y.
\end{aligned}
\end{equation}
Thus, compactness, charge and the surface-to-scale ratio cannot be specified independently. At fixed $y$,
\begin{equation}
\left.\frac{\partial u}{\partial\varepsilon}\right|_y
=\frac{y^2(1+2y)}{1+3y}>0,
\end{equation}
so increasing the charge parameter increases the compactness of the matched solution. It also changes $\beta$, and therefore the dimensional scale when the physical bag constant is held fixed. The deformation coupling $\alpha$ introduced earlier is not an additional independent parameter of this reconstructed family.

To compare configurations at equivalent positions within their allowed charge intervals, we define
\begin{equation}
f=\frac{\varepsilon}{\varepsilon_{\rm max}(y)},
\end{equation}
where $\varepsilon_{\rm max}(y)$ is the surface tangential-pressure bound in Eq.~\eqref{eq:generalbound}. This definition is used for $y>y_0$, where the bound is positive. A fixed value of $f$ represents the same fraction of the permitted charge amplitude, rather than the same total charge or electric field across different geometries.

Our selection requires a positive bag parameter, regular positive metric potentials, finite central material variables, central pressure isotropy, and nonnegative, nonincreasing material density and pressures. The null, weak, dominant and strong energy conditions are evaluated for both the material and total sources. Matching and hydrostatic balance are checked separately. The exterior must also remain static for $r\geq R$; when $M^2\geq Q^2$, this requires
$R>M+\sqrt{M^2-Q^2}$.
These conditions define the equilibrium class considered here. In particular, nonnegative anisotropy is not imposed: a negative value of $p_t-p_r$ represents an inward anisotropic contribution to the force balance and is not, by itself, an energy-condition violation \cite{bowers}.

For a diagonal source with energy density $\varrho$ and principal pressures $P_r$ and $P_t$, the tested inequalities are
\begin{align}
\mathrm{NEC}:&\quad
\varrho+P_r\geq0,\quad \varrho+P_t\geq0,\\
\mathrm{WEC}:&\quad
\mathrm{NEC},\quad \varrho\geq0,\\
\mathrm{DEC}:&\quad
\varrho-|P_r|\geq0,\quad \varrho-|P_t|\geq0,\\
\mathrm{SEC}:&\quad
\mathrm{NEC},\quad \varrho+P_r+2P_t\geq0.
\end{align}
In the dimensionless variables used for the figures, the total source is
\begin{equation}
\begin{aligned}
\widehat{\rho}_{\rm tot}&=\widehat{\rho}+\varepsilon t,\\
\widehat{P}_{r,\rm tot}&=\widehat{p}_r-\varepsilon t,\\
\widehat{P}_{t,\rm tot}&=\widehat{p}_t+\varepsilon t.
\end{aligned}
\end{equation}
The total radial pressure is consequently
$-\varepsilon y$ at the surface. This electromagnetic tension is consistent with matching to the charged exterior; the vanishing boundary quantity is the material radial pressure.

\subsection{Global static admissibility domain}
\label{sec:globaldomain}

let $S(y)=2y^3+6y^2+3y-5$, with positive root $y_0$.

\noindent\textbf{Proposition (static admissibility).} Within the regular matched family defined by Eqs.~\eqref{eq:h} and \eqref{eq:generalbeta}, the static requirements specified above hold if and only if
\begin{equation}
\displaystyle y\geq y_0,\qquad
0\leq\varepsilon\leq\varepsilon_{\rm max}(y)
=\frac{21S(y)}{J(y)}.
\label{eq:globaldomain}
\end{equation}
The metric potentials and density are positive; both pressures are positive inside and nonnegative at $R$. Density and pressures decrease strictly for $0<r\leq R$. The material and total sources satisfy the standard energy conditions.

\noindent\textit{Proof.} Necessity follows from Eq.~\eqref{eq:generalpt}: its denominator is positive for $y>0$ and $S$ has one positive root. At fixed $y$, the matched $\beta$ and the expressions in Table~\ref{tab:certificateconstruction} are affine in $\varepsilon$. Thus we need only check $\varepsilon=0$ and $\varepsilon=\varepsilon_{\rm max}(y)$. At $y=y_0$ they coincide.

In the Bernstein basis on $\zeta=t/y\in[0,1]$ (Appendix~\ref{app:certificates}), the first nine rows of Table~\ref{tab:certificateconstruction} have nonnegative coefficients for $y\geq y_0$. Their initial coefficients are positive, and all coefficients outside the pressure rows are strictly positive. This proves positivity of the interior profiles, monotonicity of the material variables and the dominant energy inequalities. Also, $H_t=V/(1+t)^4$, where $-V$ has nonnegative coefficients in $t$, $\beta$ and $\varepsilon$ and a unit constant term. Hence $H_t<0$. The density, pressure and dominant-energy inequalities imply the standard material energy conditions. Since the Maxwell tensor also satisfies them, so does the total source.

The surface also satisfies $\varepsilon y^2<1$, since
\begin{equation}
\begin{aligned}
J-21y^2 S={}&498y^5+2536y^4+5137y^3\\
&+5082y^2+2261y+350>0.
\end{aligned}
\end{equation}
Hence $b^2>0$, $F_R>0$ and $M/R<1$. For $M\geq|Q|$, $R$ lies outside the outer Reissner--Nordstr\"om horizon; otherwise no horizon exists. Regularity follows from the central expansions and cubic charge law. Matching and hydrostatic balance hold identically. This completes the proof.

\subsection{Tangential equilibrium-profile derivative}

Both profiles decrease, so $dp_t/d\rho>0$. The upper bound follows from the last row of Table~\ref{tab:certificateconstruction}. At maximum charge, only its final coefficient can change sign:
\begin{equation}
\begin{aligned}
c_6={}&\frac{6(1+y)^5}{J(y)}W(y),\\
W(y)={}&-2y^4+136y^3+693y^2\\
&+1190y+455.
\end{aligned}
\label{eq:tangentialbernsteinlimit}
\end{equation}
By Descartes' rule and the signs at zero and infinity, $W$ has a unique positive root, $y_{\rm crit}\simeq72.86783136$. For
\begin{equation}
y_0\leq y<y_{\rm crit},\quad
0\leq\varepsilon\leq\varepsilon_{\rm max}(y),
\label{eq:tangentialdomain}
\end{equation}
the ratio lies in $(0,1)$, including its central limit. At maximum charge it reaches unity at the surface when $y=y_{\rm crit}$ and exceeds unity there for larger $y$. Static admissibility is unaffected, but the derivative bound would restrict the charge further. This ratio is not a characteristic speed.

\subsection{Representative families and numerical survey}

We illustrate the results for $y=3/4,1,3/2,2,3,4$, all within Eqs.~\eqref{eq:globaldomain} and \eqref{eq:tangentialdomain}.

\begin{table}[t]
\centering
\small
\caption{Continuously certified load intervals for six fixed values of $y$. The subscripts $0$ and $1$ denote $f=0$ and $f=1$, respectively. Each endpoint of the load interval is given exactly, with its decimal approximation below; other values are rounded. The certificates can be reproduced using Appendix~\ref{app:certificates}.}
\label{tab:families}
\begin{tabular}{rrrrr}
\toprule
$y$ & $\varepsilon_{\rm max}$ & $u_0$ & $u_1$ & $(Q/R)_1$\\
\midrule
0.75 & \shortstack{$\rev{\frac{7896}{2050427}}$\\[2pt]0.0038509} & 0.23077 & 0.23244 & 0.04654\\
1    & \shortstack{$\rev{\frac{21}{2665}}$\\[2pt]0.0078799} & 0.25000 & 0.25591 & 0.08877\\
1.5  & \shortstack{$\rev{\frac{1659}{200267}}$\\[2pt]0.0082839} & 0.27273 & 0.28628 & 0.13652\\
2    & \shortstack{$\rev{\frac{41}{6012}}$\\[2pt]0.0068197} & 0.28571 & 0.30520 & 0.16516\\
3    & \shortstack{$\rev{\frac{21}{4814}}$\\[2pt]0.0043623} & 0.30000 & 0.32748 & 0.19814\\
4    & \shortstack{$\rev{\frac{1617}{552086}}$\\[2pt]0.0029289} & 0.30769 & 0.34014 & 0.21648\\
\bottomrule
\end{tabular}
\end{table}

All six families satisfy the strict tangential derivative bound; their charge limits and exterior parameters appear in Table~\ref{tab:families}.

A broader numerical survey was performed on 141 equally spaced values of $y\in[0.5,4]$ and 61 values of $\varepsilon\in[0,0.03]$. Each of the resulting 8,601 configurations was evaluated at 1,001 equally spaced points in $r/R$, including the centre and surface, using analytic derivatives and a sign tolerance of $10^{-10}$. Of these configurations, 1,586 pass all static filters, and all accepted cases also satisfy the sampled tangential derivative bound.

Within this particular grid, acceptance coincides with the necessary surface restriction in Eq.~\eqref{eq:generalbound}. Every rejected configuration fails material tangential-pressure positivity; 6,275 fail this condition alone, while the remaining rejected cases also violate a strong-energy inequality. Thus, the surface stress is an effective indicator of the numerical boundary in the explored domain. The numerical survey thus agrees with Eq.~\eqref{eq:globaldomain}.

\begin{figure}[!tbp]
\centering
\includegraphics[width=\columnwidth]{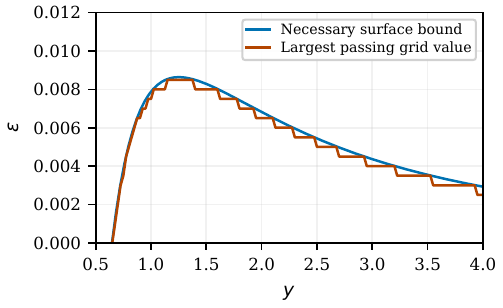}
\caption{Necessary upper charge bound from the surface
tangential pressure and the largest accepted charge on the
numerical grid. The charge spacing is
$\Delta\varepsilon=0.0005$; the stepped appearance reflects
this finite resolution.}
\label{fig:domain}
\end{figure}

\begin{figure}[!tbp]
\centering
\includegraphics[width=\columnwidth]{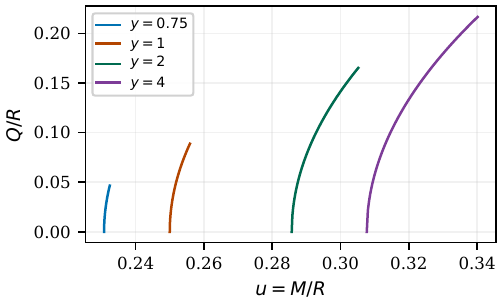}
\caption{Compactness and surface charge along four certified
fixed-$y$ families as $f$ varies from zero to unity.
The curves follow the matching relations and do not
represent independently prescribed values of $M/R$ and $Q/R$.}
\label{fig:domain-2}
\end{figure}

Figures~\ref{fig:domain} and \ref{fig:domain-2} show that the interpretation of the charge limit depends on the quantity being compared. Although $\varepsilon_{\rm max}$ decreases between $y=1.5$ and $y=4$, the largest permitted $Q/R$ in Table~\ref{tab:families} increases over the same range. The difference comes from the geometric factor in $Q/R=\sqrt{\varepsilon}\,y$: its growth can more than compensate for a reduction in the maximum allowed $\varepsilon$.

\begin{figure}[!tbp]
\centering
\includegraphics[width=\columnwidth]{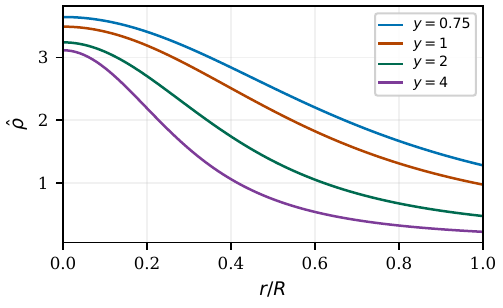}
\caption{Normalized material density for $y=0.75,1,2,4$
at $f=1/2$. The normalization uses the matched value of $a$ for each configuration. At fixed physical bag constant, the dimensional densities need not follow the same ordering.}
\label{fig:geomatter}
\end{figure}

\begin{figure}[!tbp]
\centering
\includegraphics[width=\columnwidth]{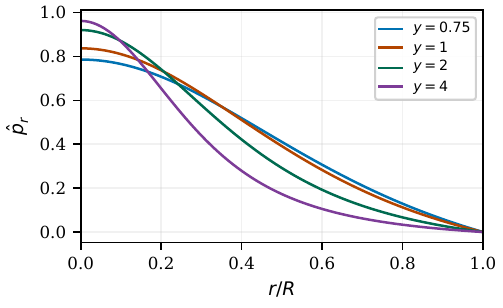}
\caption{Normalized radial material pressure at $f=1/2$.
The central pressure and radial concentration vary with
$y$, while every curve satisfies $p_r(R)=0$.}
\label{fig:geomatter-2}
\end{figure}

\begin{figure}[!tbp]
\centering
\includegraphics[width=\columnwidth]{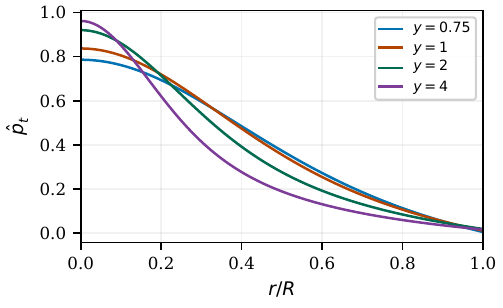}
\caption{Normalized tangential material pressure at
$f=1/2$. The profiles decrease outward and retain a
positive surface value on these matched configurations.}
\label{fig:geomatter-3}
\end{figure}

\begin{figure}[!tbp]
\centering
\includegraphics[width=\columnwidth]{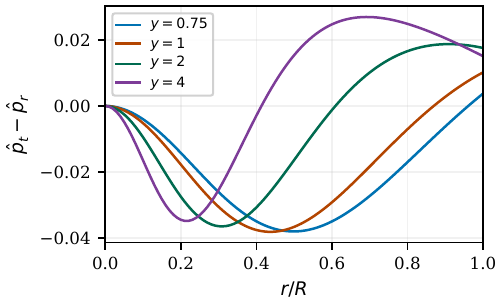}
\caption{Material pressure anisotropy at $f=1/2$.
An inward anisotropic contribution near the centre is
followed by an outward contribution in the outer region.}
\label{fig:geomatter-4}
\end{figure}

We now examine the radial profiles of these families. Figures~\ref{fig:geomatter}--\ref{fig:geomatter-4} compare $y=0.75,1,2,4$ at $f=1/2$. All four configurations have finite central values and decreasing density and pressures. The radial pressure reaches zero at the boundary, whereas the tangential pressure remains positive because the charge lies below the limiting value.

The normalization is important when interpreting the density curves. At a fixed physical $\mathcal{B}$,
\begin{equation}
\rho=\mathcal{B}\frac{\widehat{\rho}}{\beta},
\qquad
p_i=\mathcal{B}\frac{\widehat{p}_i}{\beta}.
\end{equation}
The central and surface relations therefore give
\begin{equation}
\begin{aligned}
\frac{\rho_c}{\mathcal{B}}&=\frac{3}{\beta}+2,\\
\frac{p_{r,c}}{\mathcal{B}}&=\frac{1}{\beta}-\frac{2}{3},\\
\frac{\rho_c}{\rho_s}&=\frac{3+2\beta}{4\beta}.
\end{aligned}
\end{equation}
For the four configurations shown, $\rho_c/\rho_s$ increases from approximately $2.84$ at $y=0.75$ to $13.71$ at $y=4$. Thus, the lower normalized density curve at larger $y$ accompanies a greater central-to-surface density contrast. At fixed $\mathcal{B}$, the central physical density also increases across these examples. The figures describe increasingly concentrated configurations, rather than a simple decrease of physical density with compactness.

The sign change of the anisotropy can also be understood analytically. Expansion of the reconstructed pressures gives
\begin{equation}
\widehat{\Delta}
\equiv\widehat{p}_t-\widehat{p}_r
=-\frac{3+\beta+15\varepsilon}{6}\,t + \mathcal{O}(t^2).
\label{eq:anisotropycentre}
\end{equation}
Since $\beta>0$ and $\varepsilon\geq0$, the material anisotropy is necessarily negative sufficiently close to the centre. For $f<1$, its surface value is positive, because $p_r(R)=0$ and $p_t(R)>0$. A transition between inward and outward anisotropic contributions is therefore required. In the sampled $f=1/2$ profiles, this transition moves from approximately $r/R=0.96$ at $y=0.75$ to $r/R=0.41$ at $y=4$.

The anisotropy in the interior region, expressed in Eq.~\eqref{eq:anisotropycentre}, is compatible with positive material pressures and the established energy conditions. Here $p_{r}-p_{t}$ does not imply a negative material pressure. The mechanisms discussed in the Introduction offer potential material interpretations for such a stress configuration; however, the temporal ansatz and the radial closure employed here determine only its effective macroscopic profile. The inward-directed contribution, $2\Delta/r$, must be compensated by the other terms in Eq.~\eqref{eq:tov}; its sign, by itself, does not imply dynamical instability. In fact, models of quark stars with interactions allow for radially stable configurations with negative anisotropy, given specific constitutive and perturbative prescriptions \cite{pretel2024interacting}.

This behaviour is relevant when comparing the solution with other charged anisotropic bag models. Previous constructions have examined the competing roles of anisotropic and electric forces, including a change in their relative importance between the inner and outer regions \cite{maurya}. Here the material anisotropy itself changes direction. The anisotropic force therefore does not oppose gravity throughout the star.

\begin{figure}[!tbp]
\centering
\includegraphics[width=\columnwidth]{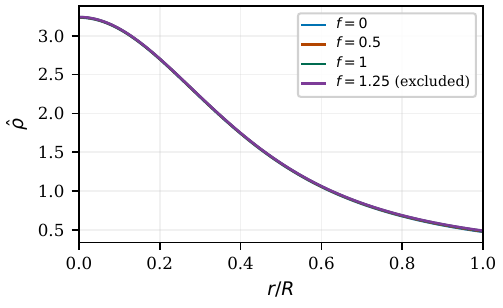}
\caption{Normalized material density at $y=2$ for four
charge amplitudes. The curves are close despite their
different tangential-pressure behaviour. The $f=1.25$
configuration is excluded by the adopted pressure criterion.}
\label{fig:chargematter}
\end{figure}

\begin{figure}[!tbp]
\centering
\includegraphics[width=\columnwidth]{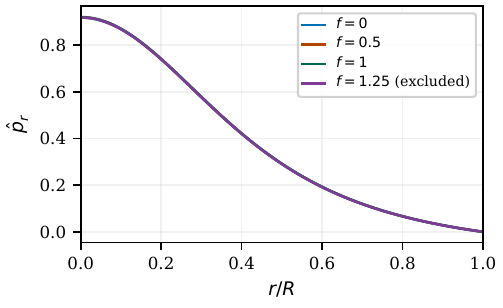}
\caption{Normalized radial pressure at $y=2$. All four curves have $p_r(R)=0$ by construction, even for the configuration excluded by the pressure criterion.}
\label{fig:chargematter-2}
\end{figure}

\begin{figure}[!tbp]
\centering
\includegraphics[width=\columnwidth]{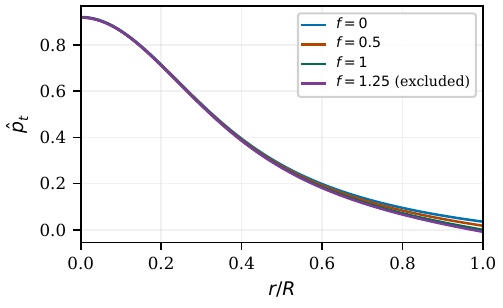}
\caption{Normalized tangential pressure at $y=2$.
The surface value decreases with charge, vanishes at
$f=1$, and becomes negative for $f=1.25$.}
\label{fig:chargematter-3}
\label{fig:electric-4}
\end{figure}

\begin{figure}[!tbp]
\centering
\includegraphics[width=\columnwidth]{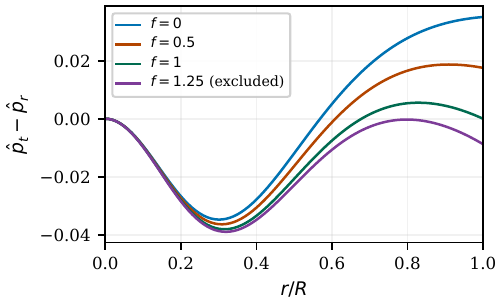}
\caption{Material anisotropy at $y=2$. Increasing charge
changes the extent of the inward and outward anisotropic
regions and reduces the surface anisotropy.}
\label{fig:chargematter-4}
\end{figure}

Figures~\ref{fig:chargematter}--\ref{fig:chargematter-4} compare $f=0,1/2,1,5/4$ at $y=2$. The density and radial-pressure curves nearly overlap, whereas the tangential pressure and anisotropy respond more noticeably to charge. The increase in $\beta$ with $\varepsilon$ at fixed $y$ accounts for the slight rise in $\widehat{\rho}_c=3+2\beta$ and the reduction in $\widehat{p}_{r,c}=1-2\beta/3$. These are statements about normalized variables. At fixed physical bag constant, $\rho_c/\mathcal{B}=3/\beta+2$ instead decreases as charge increases.

The surface response has a particularly simple form. At fixed $y$, Eq.~\eqref{eq:generalpt} implies
\begin{equation}
\widehat{p}_t(R;f)
=(1-f)\widehat{p}_t(R;0).
\end{equation}
For $y=2$, the neutral surface value is approximately
$0.03507271$, decreasing to $0.01753636$ at $f=1/2$ and zero at $f=1$. At $f=5/4$, it becomes approximately $-0.00876818$. The electric contribution to the total tangential stress remains positive, but this does not prevent the material tangential pressure from becoming negative. The exclusion therefore concerns the adopted material model, not the absence of a formal Einstein--Maxwell solution.

\begin{figure}[!tbp]
\centering
\includegraphics[width=\columnwidth]{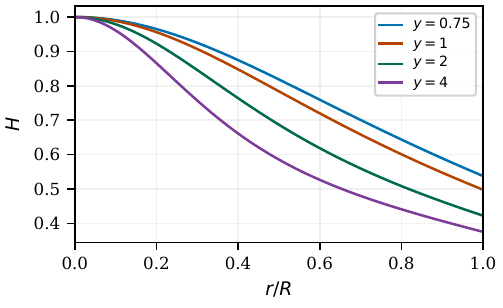}
\caption{Radial metric potential at $f=1/2$. All profiles
start from $H(0)=1$ and remain positive throughout the
interior.}
\label{fig:geometry}
\end{figure}

\begin{figure}[!tbp]
\centering
\includegraphics[width=\columnwidth]{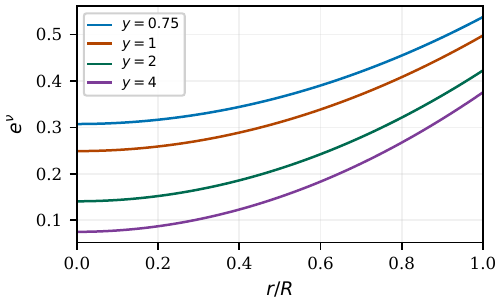}
\caption{Temporal metric potential at $f=1/2$, with its
normalization fixed by matching to the exterior time
coordinate.}
\label{fig:geometry-2}
\end{figure}

\begin{figure}[!tbp]
\centering
\includegraphics[width=\columnwidth]{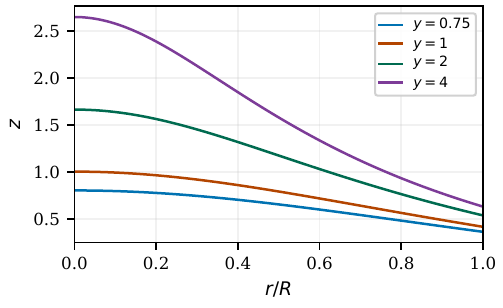}
\caption{Gravitational redshift to infinity at $f=1/2$.
The larger-$y$ configurations exhibit a greater contrast
between central and surface redshifts.}
\label{fig:geometry-3}
\end{figure}

\begin{figure}[!tbp]
\centering
\includegraphics[width=\columnwidth]{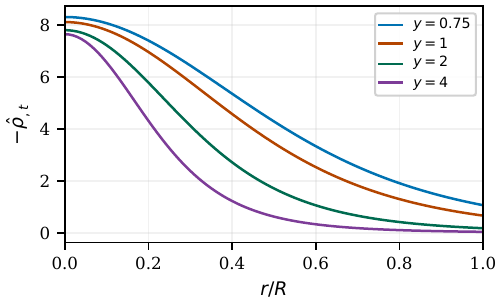}
\caption{Signed density derivative at $f=1/2$.
Positive $-\widehat{\rho}_{,t}$ establishes decreasing
density for $r>0$; the radial derivative vanishes at the
centre by regularity.}
\label{fig:geometry-4}
\end{figure}

The metric profiles in Figures~\ref{fig:geometry}--\ref{fig:geometry-4} remain regular and positive. The two metric potentials have opposite radial trends: the temporal component grows with radius, whereas $H$ falls from its central value of unity. The normalization obtained at the boundary determines the redshift of a static emitter at any interior radius. For $f=1/2$, the surface redshift increases from about $0.364$ at $y=0.75$ to $0.632$ at $y=4$; the corresponding central values rise from approximately $0.804$ to $2.649$. The greater redshift contrast accompanies the increased density concentration. The central redshift describes a hypothetical static emitter inside the configuration and is not an observable surface redshift.

\begin{figure}[!tbp]
\centering
\includegraphics[width=\columnwidth]{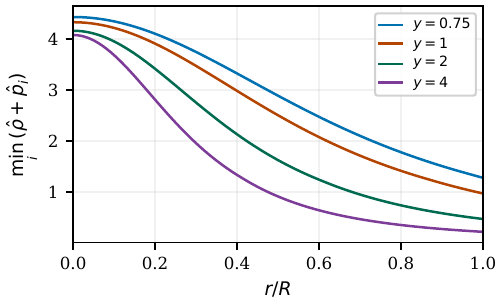}
\caption{Minimum material NEC margin at $f=1/2$.
Together with positive density, these inequalities also
establish the material WEC.}
\label{fig:energy}
\end{figure}

\begin{figure}[!tbp]
\centering
\includegraphics[width=\columnwidth]{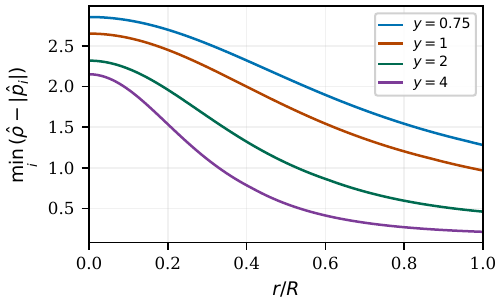}
\caption{Minimum material DEC margin at $f=1/2$.
The material density exceeds the magnitude of both
principal pressures throughout the plotted configurations.}
\label{fig:energy-2}
\end{figure}

\begin{figure}[!tbp]
\centering
\includegraphics[width=\columnwidth]{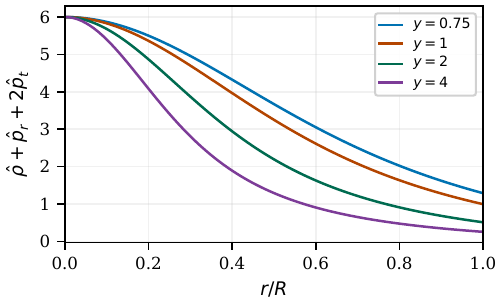}
\caption{Material combination
$\widehat{\rho}+\widehat{p}_r+2\widehat{p}_t$
at $f=1/2$. Its positivity, together with the NEC,
establishes the material SEC.}
\label{fig:energy-3}
\end{figure}

\begin{figure}[!tbp]
\centering
\includegraphics[width=\columnwidth]{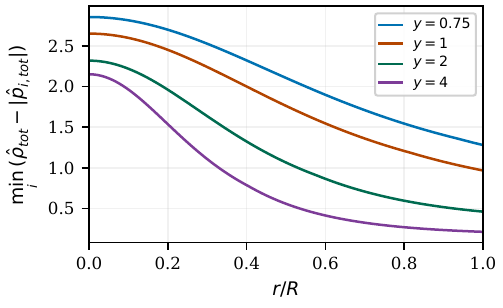}
\caption{Minimum total DEC margin at $f=1/2$, including
the electromagnetic energy density and principal stresses.}
\label{fig:energy-4}
\end{figure}

The energy-condition margins remain positive for these configurations, as shown in Figures~\ref{fig:energy}--\ref{fig:energy-4}. Their behaviour should be interpreted together with the material-pressure test. In particular, the excluded $y=2$, $f=5/4$ configuration still satisfies all sampled material and total energy inequalities, with a smallest recorded energy-condition margin of approximately $0.4693$. Energy conditions constrain combinations of density and stress and can allow a limited tension. They therefore do not replace the separate requirement of nonnegative material pressures adopted in this work.

The relation between the material and total tests can be followed directly. The electric field leaves the radial NEC combination unchanged, while adding $2\varepsilon t$ to the tangential NEC and to
$\widehat{\rho}+\widehat{p}_r+2\widehat{p}_t$.
For nonnegative material tangential pressure, the total tangential DEC margin is exactly
$\widehat{\rho}-\widehat{p}_t$, because the electromagnetic contributions cancel. The total radial DEC follows from
\begin{equation}
\begin{aligned}
\widehat{\rho}+\varepsilon t
-\left|\widehat{p}_r-\varepsilon t\right|
\geq \widehat{\rho}-|\widehat{p}_r|.
\end{aligned}
\end{equation}
The material and total energy plots are consequently related checks, rather than independent evidence of stability.

\begin{figure}[!tbp]
\centering
\includegraphics[width=\columnwidth]{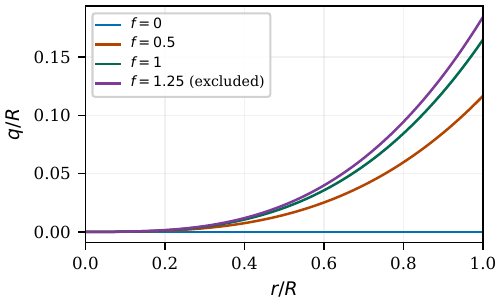}
\caption{Enclosed charge at $y=2$. The cubic radial
dependence is common to all charged curves, while the
amplitude is proportional to $\sqrt{\varepsilon}$.}
\label{fig:electric}
\end{figure}

\begin{figure}[!tbp]
\centering
\includegraphics[width=\columnwidth]{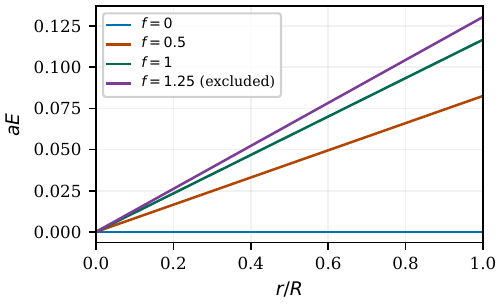}
\caption{Normalized electric field at $y=2$. Its linear
radial dependence ensures a vanishing central field.
The excluded curve retains a regular electric profile.}
\label{fig:electric-2}
\end{figure}

\begin{figure}[!tbp]
\centering
\includegraphics[width=\columnwidth]{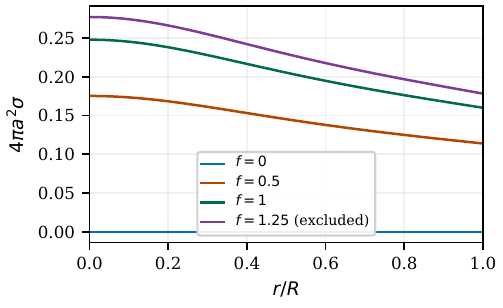}
\caption{Normalized proper charge density at $y=2$.
The factor $\sqrt{H}$ produces a nonuniform density
despite the simple cubic law for enclosed charge.}
\label{fig:electric-3}
\end{figure}

The electrical profiles in Figures~\ref{fig:electric}--\ref{fig:electric-3} display the regular cubic charge prescription. With $\xi=r/R$,
\begin{equation}
\begin{aligned}
\frac{q(r)}{R}
&=\sqrt{\varepsilon}\,y\,\xi^3,\\
aE(r)&=\sqrt{\varepsilon y}\,\xi,\\
4\pi a^2\sigma(r)&=3\sqrt{\varepsilon}\sqrt{H(r)}.
\end{aligned}
\end{equation}
Both the enclosed charge and electric field vanish at the centre and increase outward, while the proper charge density decreases with $H$. Thus, an outwardly increasing enclosed charge does not imply an outwardly increasing local charge density. The cubic prescription is shared with earlier charged strange-star studies \cite{deb2018}; here its interpretation in terms of proper charge density remains explicitly tied to the reconstructed geometry.

The force balance makes the role of these profiles more explicit. Let $F_i$ denote the dimensionless force-density terms obtained by multiplying the physical equilibrium terms by $8\pi a^3$. Their expressions are
\begin{equation}
\begin{aligned}
F_p&=-2\sqrt{t}\,\widehat{p}_{r,t},\\
F_g&=-(\widehat{\rho}+\widehat{p}_r)
       \frac{\sqrt{t}}{1+t},\\
F_a&=\frac{2(\widehat{p}_t-\widehat{p}_r)}{\sqrt{t}},\\
F_e&=6\varepsilon\sqrt{t},
\end{aligned}
\end{equation}
and the equilibrium equation requires
$F_p+F_g+F_a+F_e=0$.
All four terms vanish at the centre; the apparent division by $\sqrt{t}$ in $F_a$ is regular because the anisotropy is $\mathcal{O}(t)$.

\begin{figure}[!tbp]
\centering
\includegraphics[width=\columnwidth]{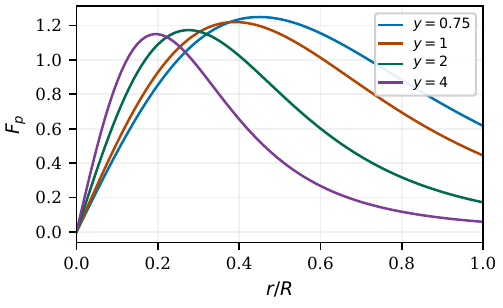}
\caption{Normalized pressure-gradient contribution at
$f=1/2$. It is outward because the radial material pressure
decreases with radius.}
\label{fig:forces}
\end{figure}

\begin{figure}[!tbp]
\centering
\includegraphics[width=\columnwidth]{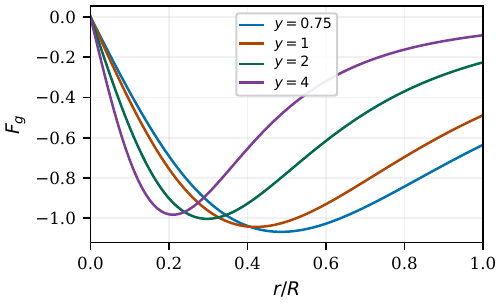}
\caption{Normalized gravitational contribution at
$f=1/2$. It remains inward throughout the interior.}
\label{fig:forces-2}
\end{figure}

\begin{figure}[!tbp]
\centering
\includegraphics[width=\columnwidth]{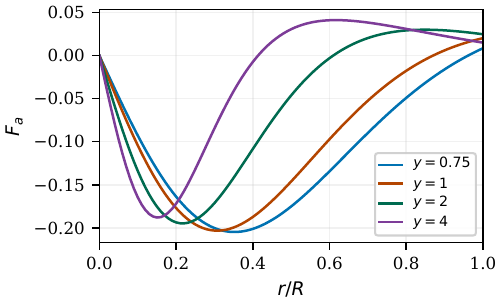}
\caption{Normalized anisotropic contribution at $f=1/2$.
Its change of sign follows the change of sign of the
material pressure anisotropy.}
\label{fig:forces-3}
\end{figure}

\begin{figure}[!tbp]
\centering
\includegraphics[width=\columnwidth]{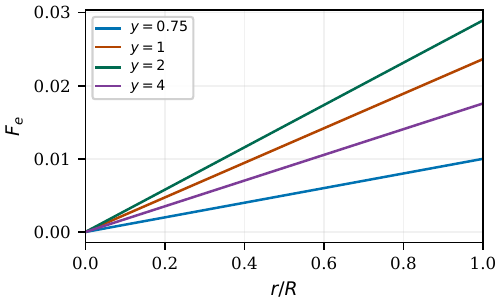}
\caption{Normalized electric force for $f=1/2$. Note that, for a given $r/R$, the electric force is proportional to $\varepsilon\sqrt{y}$, although $f$ is the same for all configurations.}
\label{fig:forces-4}
\end{figure}

Figures~\ref{fig:forces}--\ref{fig:forces-4} show that the pressure gradient balances gravity together with an inward anisotropic contribution in the central region. Farther out, the anisotropic contribution becomes outward and acts in the same direction as the electric term. The latter increases linearly with radius for the adopted charge law. This redistribution of support is consistent with the importance of electric and anisotropic forces found in other charged bag-model constructions \cite{maurya}, but the relative magnitudes are specific to the geometry and constitutive assumptions used here. In particular, a larger outward force in an equilibrium equation does not demonstrate a restoring response to a perturbation.

\begin{figure}[!tbp]
\centering
\includegraphics[width=\columnwidth]{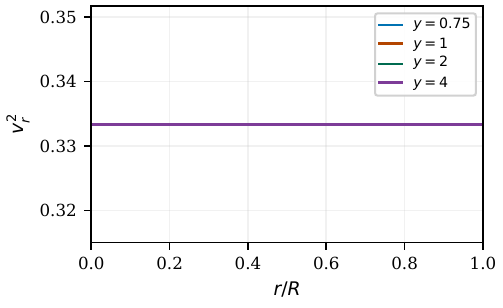}
\caption{Radial pressure--density derivative at $f=1/2$.
All four curves coincide at $1/3$, as required by the
radial bag relation.}
\label{fig:diagnostics}
\end{figure}

\begin{figure}[!tbp]
\centering
\includegraphics[width=\columnwidth]{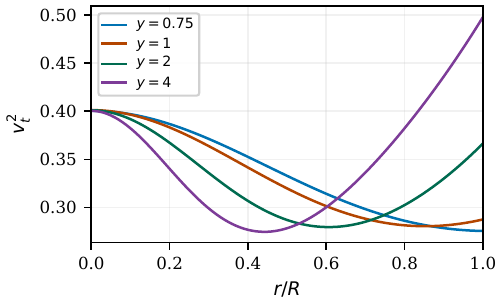}
\caption{Tangential equilibrium-profile derivative at
$f=1/2$. The larger-$y$ examples develop an outward rise
after an interior minimum while remaining below unity.}
\label{fig:diagnostics-2}
\end{figure}

\begin{figure}[!tbp]
\centering
\includegraphics[width=\columnwidth]{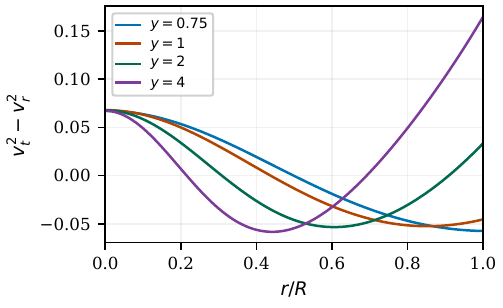}
\caption{Difference between tangential and radial profile
derivatives at $f=1/2$. The sign varies with radius and
does not provide a uniform fixed-sign classification.}
\label{fig:diagnostics-3}
\end{figure}

\begin{figure}[!tbp]
\centering
\includegraphics[width=\columnwidth]{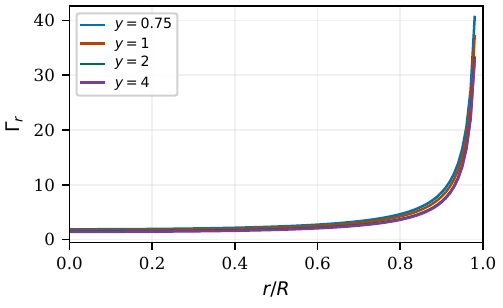}
\caption{Radial adiabatic-index expression at $f=1/2$.
The curves stop at $r/R=0.98$ because the index diverges
as the material radial pressure vanishes at the surface.}
\label{fig:diagnostics-4}
\end{figure}

The derivative diagnostics reinforce this distinction.
Figures~\ref{fig:diagnostics}--\ref{fig:diagnostics-4}
retain the conventional notation $v_r^2$ and $v_t^2$ for comparison with the literature, with
\[
v_r^2=\frac{1}{3},
\qquad
v_t^2=\frac{\widehat{p}_{t,t}}{\widehat{\rho}_{,t}}.
\]
The second quantity is an equilibrium-profile derivative. Its identification with a tangential sound speed requires a dynamical material prescription. The central value follows directly from the regular series:
\begin{equation}
v_t^2(0)=
\frac{6+2\beta+4\varepsilon}
     {15+5\beta-3\varepsilon}.
\end{equation}
It equals $2/5$ in the neutral limit and is slightly larger in the charged examples. Thus, $v_t^2-v_r^2$ is positive near the centre even though both ratios remain below unity.

The connection between differences of sound speeds and potential cracking regions has been investigated under specified perturbative assumptions \cite{abreu2007}, including applications to decoupled stellar models \cite{andrade_cracking2022}. For the curves shown, the derivative difference is not negative throughout the star. More importantly, the static relation between $p_t$ and $\rho$ does not specify how anisotropy responds away from equilibrium. These plots therefore cannot establish either the absence of cracking or the sign of a radial-mode eigenvalue \cite{horvat,pretel}.

For the radial adiabatic-index expression,
\begin{equation}
\Gamma_r=\frac{\rho+p_r}{3p_r}
=\frac{4}{3}+\frac{4\mathcal{B}}{3p_r},
\end{equation}
the excess above $4/3$ follows algebraically from the bag relation whenever $p_r>0$. Its divergence at the surface is likewise a consequence of the finite zero-pressure density. Across the displayed $f=1/2$ configurations, the central value decreases from approximately $1.878$ at $y=0.75$ to $1.412$ at $y=4$. Neither the inequality $\Gamma_r>4/3$ nor its outward increase supplies an independent demonstration of relativistic stability.

The excluded charge values provide a useful check on the interpretation of these diagnostics. For each of the six selected geometries, $f=5/4$ violates tangential-pressure positivity while retaining sampled derivative ratios between zero and one. The classification of configurations is unchanged on radial grids of 1,001, 2,001, 4,001 and 8,001 points for the 24 selected cases. This supports the numerical resolution of the profiles, while the exact certificates establish the continuous inequalities for $f\leq1$.

Additional consistency checks give maximum normalized EOS and force-balance residuals of approximately
$2.0\times10^{-15}$ and $1.1\times10^{-15}$ across the rectangular survey. These are algebraic floating-point checks. A separate DOP853 integration of the metric equation at $y=0.75,1,2,4$ and $f=1/2$ agrees with the closed-form solution within $1.4\times10^{-10}$, using relative and absolute tolerances of $10^{-12}$ and $10^{-14}$. The independent integration complements the exact identities and coefficient certificates.

The results can finally be placed on a dimensional scale by fixing the physical bag constant. For each matched pair $(y,\varepsilon)$,
\[
a=\sqrt{\frac{\beta}{8\pi\mathcal{B}}},
\qquad
R=a\sqrt{y},
\qquad
M=uR
\]
in geometrized units. Thus, varying $\mathcal{B}$ at a fixed dimensionless configuration rescales both mass and radius as $\mathcal{B}^{-1/2}$, leaving compactness and the dimensionless profiles unchanged. Table~\ref{tab:scalesmulti} illustrates this dependence.

\begin{table}[t]
\centering
\small
\caption{Illustrative dimensional scales at $f=1/2$.
The bag energy density is in MeV fm$^{-3}$, the radius
in km, and the mass in solar masses. These are matched
equilibrium configurations, not observational fits or
maximum stable masses.}
\label{tab:scalesmulti}
\begin{tabular}{rrrrr}
\toprule
$y$ & $\mathcal{B}$ & $R$ & $M/M_\odot$ & $u$\\
\midrule
0.75 & 60 & 10.980 & 1.722 & 0.23160\\
0.75 & 80 & 9.509  & 1.491 & 0.23160\\
1    & 60 & 11.063 & 1.895 & 0.25295\\
1    & 80 & 9.581  & 1.641 & 0.25295\\
2    & 60 & 10.942 & 2.189 & 0.29546\\
2    & 80 & 9.476  & 1.896 & 0.29546\\
4    & 60 & 10.667 & 2.340 & 0.32391\\
4    & 80 & 9.238  & 2.026 & 0.32391\\
\bottomrule
\end{tabular}
\end{table}

At $\mathcal{B}=60\,\mathrm{MeV\,fm^{-3}}$, the displayed configurations span approximately $1.72$--$2.34\,M_\odot$ with radii near $11$ km. Increasing the bag constant to
$80\,\mathrm{MeV\,fm^{-3}}$ reduces both quantities by the factor $\sqrt{60/80}$. The increase of mass with charge at fixed $y$ and fixed $\mathcal{B}$ follows here from the simultaneous increase of $u$ and $\beta$. Related increases of equilibrium masses have been reported in charged strange-star calculations \cite{deb2018}. The quantitative values are not directly interchangeable, since the geometric closure and the quantities held fixed along the respective sequences differ.

The same distinction applies to comparisons with Tolman-IV models employing density-dependent bag parameters \cite{goswami}. In our model $\mathcal{B}$ does not vary with radius, and we calculate the tangential pressure from the coupled equations. A common temporal potential is thus insufficient grounds for identifying the two material prescriptions or their mass limits.

In particular, the masses in Table~\ref{tab:scalesmulti} are individual equilibrium values, not determinations of a maximum stable mass. Charged-fluid pulsation studies show that turning-point interpretations depend on the charge constraint imposed along a sequence \cite{chargedstability}. For anisotropic strange stars, the treatment of nonzero surface anisotropy also affects the relation between a mass extremum and the onset of radial instability \cite{anisostability}. Both charge and surface tangential pressure vary in the comparisons made here. A stability conclusion therefore requires the corresponding perturbation laws and boundary conditions.

The main result of the survey is a controlled description of how charge and compactness reshape the material distribution while preserving exact matching. The numerical boundary agrees with the exact surface-pressure bound. The figures further show that the increased compactness is accompanied by changes in density concentration, anisotropic support and the dimensional scale. These features provide explicit equilibrium backgrounds for a subsequent dynamical analysis, with the distinction between geometric construction and material response maintained explicitly throughout.

\section{Discussion and conclusions}
\label{sec:discussion}

The study of strange stars brings together two questions that are closely related but require different kinds of evidence: how quark matter can support a compact equilibrium configuration, and which assumptions allow that configuration to be represented by a tractable relativistic model. The strange-matter hypothesis and the bag description provide the physical motivation for the first question \cite{witten,farhi,chodos}, whereas exact interior solutions offer a setting in which the interplay between the equation of state, the geometry, and the boundary conditions can be examined explicitly. Once pressure anisotropy and electric charge are included, however, this interplay becomes more restrictive, since the additional degrees of freedom must be accompanied by suitable constitutive or geometric information. This issue underlies both the classical treatment of anisotropic spheres \cite{bowers} and subsequent constructions based on gravitational decoupling \cite{ovalle,santana2022,andrade2021stellar,torres2019anisotropic}. Within this context, the present work follows the consequences of a specific set of assumptions from the compatibility of the sources to the physical restrictions on the resulting equilibrium family.

The first step was to determine whether an electromagnetic field could be introduced through a minimal radial deformation while leaving the original material distribution and temporal potential unchanged. Although this prescription might appear to be a direct application of gravitational decoupling, the electromagnetic source cannot be chosen as freely as a generic anisotropic sector: its temporal and radial mixed components satisfy the identity $S^0{}_0=S^1{}_1$. Under the assumptions considered here, this identity requires the radial deformation to satisfy $g'=\nu'g$, whose solution is incompatible with a regular nonzero deformation at the centre. The resulting obstruction therefore identifies the assumptions that must be relaxed before a charged interior can be constructed. It does not exclude charged solutions obtained through gravitational decoupling, including those in which the seed already contains an electromagnetic field and the additional source is specified independently \cite{morales2018charged,albalahi,sadiq2022}. The distinction is also consistent with charged black-hole constructions that deform a Reissner--Nordstr\"om seed using a separate sector constrained by a non-local equation of state \cite{tello2024nonlocal}, although their horizon boundary conditions differ fundamentally from the central regularity required for a stellar interior.

This compatibility result leads naturally to the coupled reconstruction adopted in the remainder of the work. Rather than keeping the material profiles fixed while adding charge, the Einstein--Maxwell equations are solved with the geometry and material stresses responding together, subject to the temporal Tolman-IV potential, a constant bag parameter, and a regular cubic charge distribution. The importance of treating these quantities together can also be seen in the equilibrium equation, where the material pressure gradient balances gravitational, anisotropic, and electromagnetic contributions \cite{bekenstein}. We impose the bag relation on the radial material pressure and use the remaining Einstein equation to calculate the tangential pressure. Even when the charge vanishes, the fluid is generally anisotropic. The retained temporal metric component is that of Tolman IV, but the neutral interior should not be confused with the original isotropic solution \cite{tolman}.

Several of the ingredients in this construction have precedents in the literature on charged quark stars. Cubic charge distributions and bag equations of state have already been considered in anisotropic strange-star models \cite{deb2018}, while other constructions combine charged matter with Tolman--Kuchowicz geometry or Tolman-IV potentials and a density-dependent bag parameter \cite{maurya,goswami}. A particularly close methodological comparison is provided by charged strange-star models in which an isotropic Einstein--Maxwell seed satisfying the MIT bag relation is extended through an additional source closed by the complexity factor \cite{sadiq2022}. In the present construction, by contrast, the constant radial bag relation and the prescribed charge profile determine the radial metric, after which the material anisotropy is reconstructed. These different closure choices explain why models sharing a bag description or a temporal potential need not have the same pressure profiles or admissible parameter ranges. Our results combine the obstruction arising from the Maxwell stress identity and central regularity with exact reconstruction and admissibility bounds. Equation~\eqref{eq:materialtrace} provides the trace of the effective anisotropic source.

The surface conditions are central to this connection because they determine which parameters can actually be varied independently. Matching to the Reissner--Nordstr\"om exterior with continuous charge and no surface layer imposes $p_r(R)=0$ and fixes the temporal normalization, while the bag relation gives the finite surface density $\rho(R)=4\mathcal{B}$. This is the expected self-bound behaviour of the adopted equation of state, rather than a failure of regularity at the boundary. At the same time, the electromagnetic radial tension remains present on both sides of the surface, so vanishing material radial pressure should not be confused with vanishing total radial stress. A nonzero material tangential pressure at the boundary is likewise compatible with the stated matching conditions. Consequently, changing the charge amplitude at fixed $R/a$ also changes the dimensionless bag parameter and the exterior quantities; the resulting curves describe a readjustment of the complete equilibrium configuration.

The effect of that readjustment is particularly clear in the tangential pressure, which supplies an explicit restriction on the charge amplitude. For the branch $R=a$, the interval
\[
0\leq\varepsilon\leq\frac{21}{2665}
\]
is necessary and sufficient for nonnegative material pressures within the family considered, and the remaining stated static inequalities are supported by exact sign certificates. The extension to other values of $R/a$ shows how this restriction changes with the geometry and demonstrates that increasing the charge does not preserve all the material properties automatically. In particular, the selected configurations beyond the admissible charge range develop negative tangential pressure, even when other profiles remain smooth. This behaviour gives the bound a direct physical interpretation within our acceptance criteria: it marks the loss of nonnegative tangential material pressure, rather than a singularity of the metric or a universal upper limit on the charge of a relativistic star.

The parameter survey provides a broader view of the same mechanism. Of the $8\,601$ sampled parameter pairs, $1\,586$ satisfy the adopted static filters, with tangential-pressure positivity providing the principal restriction in the sampled domain. Equation \eqref{eq:globaldomain} extends the surface criterion to the entire region $y \geq y_0$, with a strict bound on the derivative over the full loading interval for $y < y_{\rm crit}$. An independent integration confirms the radial metric. Pressure positivity and the energy conditions are distinct requirements. The electric field adds to the energy density and tangential pressure but subtracts from the radial pressure. We check the energy conditions for each source definition, with and without the electromagnetic contribution. Beyond the permitted charge interval, some solutions still satisfy these conditions even though the material tangential pressure becomes negative. Smooth pressure profiles do not guarantee the energy inequalities either. Related charged and decoupled stellar models commonly examine these quantities together \cite{morales2018charged,albalahi,sadiq2022}; here, their separate evaluation clarifies which physical requirement excludes a configuration and avoids treating the different tests as interchangeable.

The anisotropy profile gives a complementary interpretation of the force balance. For the present solutions, the material stress difference vanishes at the centre but becomes negative in its neighbourhood, whereas configurations with positive tangential surface pressure reach a positive stress difference at the boundary because $p_r(R)=0$. Thus, the anisotropic contribution to the equilibrium equation need not act in the same direction throughout the interior. Near the centre it contributes inward when $p_t-p_r<0$, while farther out it can provide an outward contribution where the sign reverses. The electric contribution is outward for the chosen monotonic charge distribution, but its presence does not require the material anisotropy to be outward everywhere. This behaviour is consistent with the separate roles played by charge and anisotropy in the relativistic force balance \cite{bowers,bekenstein,maurya}, and explains why interpreting every increase in charge as an increase in anisotropic support would miss part of the response of the model. Therefore, negative material anisotropy is an admissible feature of these effective equilibria, consistent with the phenomenological interpretation analyzed in Ref.~\cite{campos2026}. Its microscopic origin requires a constitutive law for anisotropy.

Similar care is needed when comparing configurations with different geometric parameters. Within the matched family, increasing the charge amplitude at fixed $R/a$ raises the compactness while reducing the tangential pressure at the surface, so a larger mass-to-radius ratio can coexist with a smaller margin of material acceptability. The length scale also matters in comparing the normalized densities. For a specified physical value of $\mathcal{B}$, the matching relation fixes that scale separately for each configuration. It must be restored before comparing dimensional densities, masses or radii. The illustrative stellar scales obtained in this way demonstrate how the exact family can be expressed in physical units, but they do not constitute a fit to an observed object or establish a maximum supported mass. Comparisons with other strange-star constructions are therefore most informative when the bag prescription, charge distribution, and anisotropy closure are specified alongside the stellar quantities \cite{deb2018,maurya,goswami}.

The radial equation of state also helps separate properties built into the construction from information supplied by the acceptability analysis. With a constant bag parameter, the material derivative $dp_r/d\rho=1/3$ follows identically, and the corresponding adiabatic-index expression satisfies $\Gamma_r>4/3$ wherever $p_r>0$. These relations provide useful consistency checks, but they do not determine the tangential material response or establish the dynamical behaviour of the configuration. In particular, the ratio $p_t'/\rho'$ describes how two equilibrium profiles vary together; interpreting it as a physical propagation speed requires additional constitutive information. This distinction is relevant when comparing static diagnostics used in anisotropic stellar models \cite{andrade_cracking2022,abreu2007,torres2019anisotropic}, since the equilibrium geometry alone does not supply a complete material description away from equilibrium.

The main physical questions left open by the construction concern precisely that material description and the origin of the charge distribution. For the case $R=a$, Eq.~\eqref{eq:schwingercriterion} reduces the charge limit from an order of $10^{-2}$ to $10^{-10}$, without establishing charge retention or equilibrium within QED. The cubic charge law ensures a regular central electric field, but it does not explain charge separation, screening, or the persistence of a macroscopic charge over stellar timescales. Similarly, imposing the bag relation on the radial pressure leaves the microscopic origin of the reconstructed tangential stress unspecified. These limitations define a useful direction for extending the model: a constitutive treatment of anisotropy and a physically motivated electrical prescription could be introduced while retaining the present exact solution as a reference configuration. Other geometric or matter closures, including those used in polytropic and complexity-based decoupling constructions \cite{santana2022,andrade2021stellar,sadiq2022}, would then allow the robustness of the admissible domain to be assessed against changes in the assumptions that close the field equations.

A natural subsequent investigation is a radial-pulsation analysis of the accepted configurations, once the perturbative anisotropy response and the electrical transport assumptions have been specified. Previous studies of charged and anisotropic strange stars show why these choices, including the treatment of surface anisotropy, must enter the dynamical problem explicitly \cite{pretel,anisostability,chargedstability}. Such an analysis would determine which part of the statically admissible family also supports stable radial modes and would provide a basis for interpreting sequences of stellar masses and radii dynamically. The results of this work concern equilibrium. A regular charged interior cannot be generated by the Maxwell-only minimal deformation under the stated assumptions. Solving for the material stresses and the geometry together instead gives an exact matched family, with explicit inequalities restricting its admissibility. This family provides a background on which the effects of charge, material anisotropy and the quark-matter prescription can be studied further.

\appendix
\renewcommand{\theHequation}{appendix.\arabic{equation}}
\section{\rev{Construction of the exact sign certificates}}
\label{app:certificates}
\begin{revision}
To reproduce the certificates using rational arithmetic, first substitute the matched $\beta(y,\varepsilon)$ from Eq.~\eqref{eq:generalbeta} into $N$ in Eq.~\eqref{eq:h}, and define
\begin{equation}
\begin{aligned}
U&=N/30,&
L&=\frac{U-(1+t)^3}{t},\\
V&=(1+t)U'-3U.&
\end{aligned}
\end{equation}
The numerator in $L$ vanishes at $t=0$, so the division by $t$ leaves a polynomial. We next define
\begin{equation}
\begin{aligned}
R_\rho&=-(1+t)L-2V-\varepsilon t(1+t)^4,\\
R_r&=(1+t)L+2U+\varepsilon t(1+t)^4,\\
R_t&=(2+t)U+(1+2t)V-\varepsilon t(1+t)^5,
\end{aligned}
\label{eq:certificatepolynomials}
\end{equation}
and the polynomial derivative operator
\begin{equation}
\mathcal{D}_k P=(1+t)P'-kP.
\label{eq:certificatederivative}
\end{equation}
The operator $\mathcal{D}_k$ acts on the polynomial $P$ and does not introduce an additional material variable.
\end{revision}

\begin{table}[t]
\centering
\begin{revision}
\caption{Numerators $P_F$ and denominator powers for
$F=P_F/(1+t)^k$. Hats on material variables are omitted.
All numerator expressions are polynomials.}
\label{tab:certificateconstruction}
\small
\begin{tabular}{lll}
\toprule
$F$ & $P_F$ & $k$\\
\midrule
$H$ & $U$ & 3\\
$\rho$ & $R_\rho$ & 4\\
$p_r$ & $R_r$ & 4\\
$p_t$ & $R_t$ & 5\\
$-\rho_t$ & $-\mathcal{D}_4R_\rho$ & 5\\
$-(p_r)_t$ & $-\mathcal{D}_4R_r$ & 5\\
$-(p_t)_t$ & $-\mathcal{D}_5R_t$ & 6\\
$\rho-p_r$ & $R_\rho-R_r$ & 4\\
$\rho-p_t$ & $(1+t)R_\rho-R_t$ & 5\\
$(p_t)_t-\rho_t$ & $\mathcal{D}_5R_t-(1+t)\mathcal{D}_4R_\rho$ & 6\\
\bottomrule
\end{tabular}
\end{revision}
\end{table}

Keep $y$ symbolic and evaluate the polynomials at $\varepsilon = 0$ and $\varepsilon=\varepsilon_{\rm max}(y)$. If $P_F(t)=\sum_{j=0}^n a_jt^j$, set $\zeta=t/y$ and calculate

\begin{equation}
\begin{aligned}
P_F(y\zeta)&=\sum_{i=0}^n c_i
 \binom{n}{i}\zeta^i(1-\zeta)^{n-i},\\
c_i&=\sum_{j=0}^{i}a_jy^j
 \frac{\binom{i}{j}}{\binom{n}{j}}.
\end{aligned}
\label{eq:generalbernstein}
\end{equation}
Here $n$ is the actual polynomial degree, after dropping any zero leading coefficients. In the order of Table~\ref{tab:certificateconstruction}, the degrees are $(4,4,4,5,3,3,4,4,5,4)$ at zero charge and $(5,5,5,6,5,5,6,5,6,6)$ at maximum charge for $y>y_0$. The same upper-endpoint expansion extends to $y=y_0$ by continuity. For $y=1$, multiplying $P_F$ and all its Bernstein coefficients by $37$ or $2665$ at the respective endpoints reproduces the normalization and minima in Table~\ref{tab:certificate}.

Equation \eqref{eq:generalbernstein} provides 50 coefficients for zero load and 64 for maximum load. Three of these vanish identically: the final coefficient of $p_r$ at both ends, and the final coefficient of $p_t$ in the case of maximum load. Of the remaining 111 functions, 109 feature polynomials in the numerator and denominator that, following simplification, have positive coefficients for $y>0$. The exceptions are
\begin{eqnarray}
c_5(p_t,0) &=& \frac{3y(1+y)^4 S(y)}{(3y+1)A(y)},\\
c_6((p_t)_t-\rho_t,\varepsilon_{\rm max})
&=& \frac{6(1+y)^5 W(y)}{J(y)},
\label{eq:exceptionalcoefficients}
\end{eqnarray}
where $W$ is given by Eq.~\eqref{eq:tangentialbernsteinlimit}. The first term is non-negative for $y\geq y_{0}$, vanishing at $y_{0}$; the second is positive for $0<y<y_{\rm crit}$. The first nine rows certify static admissibility, while the last one gives rise to Eq.~\eqref{eq:tangentialdomain}. At $y=y_{\rm crit}$ and under maximum load, only the final coefficient of that row vanishes. Consequently, the derivative ratio reaches unity at $\zeta=1$, but remains below unity in the interior.

\section*{Acknowledgment of AI assistance}

The authors acknowledge the use of GPT-6.2 (OpenAI) for English-language editing and assistance in checking symbolic calculations. All mathematical derivations and scientific results were independently verified by the authors, who take full responsibility for the content of this work.

\end{document}